%% file: ms.tex
\documentclass[acmtog,nonacm]{acmart}
\acmSubmissionID{578}

\usepackage{booktabs} 

\usepackage{dblfloatfix}
\usepackage{amsmath}

\usepackage[ruled]{algorithm2e} 

\SetAlFnt{\small}
\SetAlCapFnt{\small}
\SetAlCapNameFnt{\small}
\SetAlCapHSkip{0pt}

\usepackage{xcolor}
\usepackage{colortbl}

\usepackage{multirow}
\usepackage{subcaption}
\usepackage{tabularx}
\usepackage{graphicx}
\usepackage{xr}

\setcopyright{acmlicensed}\acmJournal{TOG}
\acmYear{2022}\acmVolume{00}\acmNumber{00}\acmArticle{00}\acmMonth{00} \acmDOI{0000.0000}

\graphicspath{{"./figures/"}}
\include{sections/definitions}

\begin{document}
\title{Time-multiplexed Neural Holography:\\A Flexible Framework for Holographic Near-eye Displays with Fast Heavily-quantized Spatial Light Modulators}

\author{Suyeon Choi}
\authornote{denotes equal contribution.}
  \email{suyeon@stanford.edu}
\affiliation{%
  \institution{Stanford University}
  \country{USA}
}
\author{Manu Gopakumar}
\authornotemark[1]
\email{manugopa@stanford.edu}
\affiliation{%
  \institution{Stanford University}
  \country{USA}
}
\author{Yifan Peng}
\email{evanpeng@stanford.edu}
\affiliation{%
  \institution{Stanford University}
  \country{USA}
}
\author{Jonghyun Kim}
  \email{jonghyunk@nvidia.com}
\affiliation{%
  \institution{NVIDIA and Stanford University}
  \country{USA}
}	
\author{Matthew O'Toole}
  \email{mpotoole@cmu.edu}
\affiliation{%
  \institution{Carnegie Mellon University}
  \country{USA}
}	
\author{Gordon Wetzstein}
  \email{gordon.wetzstein@stanford.edu}
\affiliation{%
  \institution{Stanford University}
  \country{USA}
}		

\renewcommand\shortauthors{Choi, S. and Gopakumar, M. et al.}
\renewcommand{\shorttitle}{Time-multiplexed Neural Holography}

\begin{abstract}
Holographic near-eye displays offer unprecedented capabilities for virtual and augmented reality systems, including perceptually important focus cues. Although artificial intelligence--driven algorithms for computer-generated holography (CGH) have recently made much progress in improving the image quality and synthesis efficiency of holograms, these algorithms are not directly applicable to emerging phase-only spatial light modulators (SLM) that are extremely fast but offer phase control with very limited precision. The speed of these SLMs offers time multiplexing capabilities, essentially enabling partially-coherent holographic display modes. Here we report advances in camera-calibrated wave propagation models for these types of holographic near-eye displays and we develop a CGH framework that robustly optimizes the heavily quantized phase patterns of fast SLMs. Our framework is flexible in supporting runtime supervision with different types of content, including 2D and 2.5D RGBD images, 3D focal stacks, and 4D light fields. Using our framework, we demonstrate state-of-the-art results for all of these scenarios in simulation and experiment.
\end{abstract}



\begin{teaserfigure}
  \centering
	\includegraphics[width=\columnwidth]{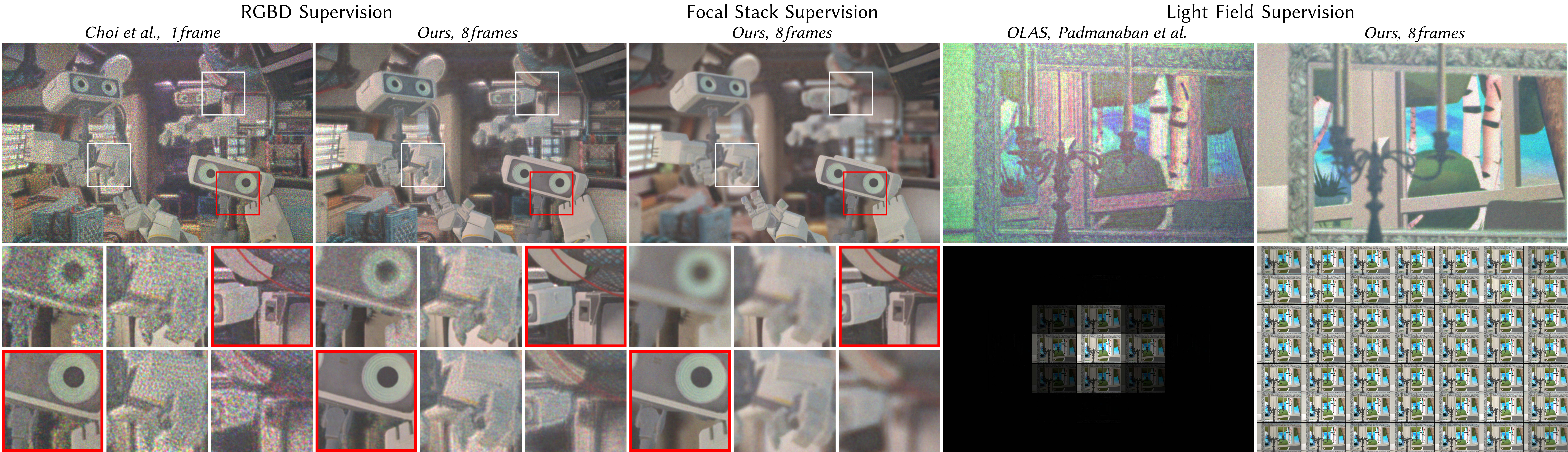}
   \caption{
	Computer-generated holography (CGH) results captured with a display prototype that uses a fast, low-precision (i.e., 4~bit) phase spatial light modulator (SLM). When supervised with 2.5D RGBD images, our approach (2nd column) provides a better image quality than the state-of-the-art neural 3D holography algorithm~\cite{choi2021neural3d} (1st column) using this low-precision SLM. Our CGH framework is flexible in not only enabling 2.5D but also 3D focal stack and 4D light field supervision. The former approach (3rd column) results in the best in-focus (red boxes) and out-of-focus (white boxes) image quality among 2.5D and 3D CGH algorithms. Our 4D light field--supervised approach (5th column) outperforms the recently proposed OLAS method~\cite{Padmanaban:2019} (4th column) by a large margin and utilizes the space--bandwidth product more effectively, as shown by the simulated light fields in the lower right images.
	}
  \label{fig:teaser}
\end{teaserfigure} 

\maketitle

\section{Introduction}
\label{sec:introduction}
\input{sections/1_introduction}

\section{Related Work}
\label{sec:related}
\input{sections/2_related}

\section{A Flexible Framework for CGH}
\label{sec:methods}
\input{sections/3_methods}

\section{Experiments}
\label{sec:experiments}
\input{sections/4_results}

\section{Discussion}
\label{sec:discussion}
\input{sections/5_discussion}

\begin{acks}
We thank Cindy Nguyen for helpful discussions. This project was in part supported by a Kwanjeong Scholarship, a Stanford SGF, Intel, NSF (award 1839974), a PECASE by the ARO (W911NF-19-1-0120), and Sony.
\end{acks}

\bibliographystyle{ACM-Reference-Format}
\bibliography{references}

\end{document}


\author{Suyeon Choi}
\authornote{denotes equal contribution.}
  \email{suyeon@stanford.edu}
\affiliation{%
  \institution{Stanford University}
  \country{USA}
}
\author{Manu Gopakumar}
\authornotemark[1]
\email{manugopa@stanford.edu}
\affiliation{%
  \institution{Stanford University}
  \country{USA}
}
\author{Yifan Peng}
\email{evanpeng@stanford.edu}
\affiliation{%
  \institution{Stanford University}
  \country{USA}
}
\author{Jonghyun Kim}
  \email{jonghyunk@nvidia.com}
\affiliation{%
  \institution{NVIDIA and Stanford University}
  \country{USA}
}	
\author{Matthew O'Toole}
  \email{mpotoole@cmu.edu}
\affiliation{%
  \institution{Carnegie Mellon University}
  \country{USA}
}	
\author{Gordon Wetzstein}
  \email{gordon.wetzstein@stanford.edu}
\affiliation{%
  \institution{Stanford University}
  \country{USA}
}		

\renewcommand\shortauthors{Choi, S. and Gopakumar, M. et al.}
\title{Time-multiplexed Neural Holography:\\A Flexible Framework for Holographic Near-eye Displays with Fast Heavily-quantized Spatial Light Modulators\Dash Supplemental Material}

\maketitle


This supplementary document includes implementation details of our holographic display prototype, complementary derivations related to wave propagation and optimization models, and additional experimental results. Refer also to the supplementary video for better visualization. \\

Here we list the abbreviations and notations used across this document. These are consistent with those in the main paper. \\

\begin{tabularx}{\columnwidth}{>{\raggedleft\arraybackslash}p{.22\columnwidth}>{\raggedright\arraybackslash}p{.72\columnwidth}}
\textbf{SLM:} & a spatial light modulator \\
\textbf{CGH:} & computer-generated holography \\
\textbf{STFT:} & the Short-time Fourier transform (STFT) \\
\textbf{ASM:} & the angular spectrum method~\cite{Goodman:2005} \\
\textbf{GS:} & the Gumber-Softmax operation~\cite{jang2016categorical,maddison2016concrete} \\
\textbf{CITL:} & the camera-in-the-loop optimization technique~\cite{peng2020neural} \\
\textbf{SGD:} & stochastic gradient descent phase retrieval~\cite{peng2020neural} \\
\end{tabularx}

\begin{tabularx}{\columnwidth}{>{\raggedleft\arraybackslash}p{.22\columnwidth}>{\raggedright\arraybackslash}p{.72\columnwidth}}
%
%
%
\textbf{NH:} & a 2D wave propagation model that is trained using an SGD-based camera-in-the-loop training strategy aka neural holography~\cite{peng2020neural}; once trained, this model is used to generate new holograms using an SGD solver \\
\textbf{NH3D:} & wave propagation model using CNNs operating on the complex-valued field at the SLM plane before ASM propagation and also directly after propagation to the target planes~\cite{choi2021neural3d} \\
\end{tabularx}


\section{Additional Details on Hardware}
\label{sec:setup}
\input{sections/suppl_setup}

\section{Additional Details on Software}
\label{sec:setup}
\input{sections/suppl_alg}

%
%
\section{Additional Experimental Results}
\label{sec:results}
\input{sections/suppl_results}

\bibliographystyle{ACM-Reference-Format}
\bibliography{references}

%% file: sections/definitions.tex
\newcommand{\loss}{\mathcal{L}}
\newcommand{\phase}{\phi}

\newcommand{\amplitude}{a}

\newcommand{\target}{\amplitude_{\text{target}}}

\newcommand{\Fourier}{\mathcal{F}}
\newcommand{\transfer}{\mathcal{H}}

\newcommand{\red}[1]{{#1}}

\usepackage[normalem]{ulem}

\newcommand{\cnn}{\textsc{cnn}} 
\newcommand{\cnnslm}{\cnn_{\textrm{\tiny SLM}}} 
 
\newcommand{\cnntarget}{\cnn_{\textrm{\tiny target}}}

\newcommand{\prop}{f}
\newcommand{\propASM}{\mathcal{P}_{\textrm{\tiny ASM}}}

\newcommand{\propModel}{\prop_{\textrm{\tiny model}}}
\newcommand{\slmAmpParam}{a_\textrm{src}}
\newcommand{\slmPhaseParam}{\phi_\textrm{src}}

\newcommand{\quant}{q}
\newcommand{\quantset}{\mathcal{Q}}
\newcommand{\gumbel}{\mathcal{G}}
\newcommand{\score}{\textbf{score}}

\DeclareRobustCommand{\thinskip}{\hskip 0.1em\relax}
\def\emdash{---}
\def\d@sh#1#2{\unskip#1\thinskip#2\thinskip\ignorespaces}
\def\Dash{\d@sh\nobreak\emdash}
\def\Ldash{\d@sh\empty{\hbox{\emdash}\nobreak}}
\def\Rdash{\d@sh\nobreak\emdash}

\newcommand{\MYhref}[3][blue]{\href{#2}{\color{#1}{#3}}}%

%% file: sections/1_introduction.tex
Holographic near-eye displays for virtual and augmented reality (VR/AR) applications offer many benefits to wearable computing systems over conventional microdisplays. These include high peak brightness, power efficiency, support of perceptually important focus cues and vision-correcting capabilities~\cite{Kim:21}, as well as thin device form factors~\cite{Maimone:2020, kim2022holographic}. Yet, the image quality achieved by computer-generated holography (CGH) lags far behind that of conventional displays, requiring further advancements in the algorithms driving holographic displays.

Recently, artificial intelligence (AI) methods have enabled significant improvements in image quality~\cite{peng2020neural,chakravarthula2020learned,choi2021neural3d} and speed~\cite{Horisaki:2018,peng2020neural,shi2021towards} of holographic displays. These algorithms, however, are primarily applicable to slow liquid crystal--based (LC) spatial light modulators (SLMs) that offer control of the phase of a coherent light source at high precision. Emerging micro-electromechanical (MEMS) phase SLMs~\cite{Bartlett:2019} offer potential benefits over LC-based systems in being more light efficient, significantly faster, better suited to operate across a wide range of wavelengths, and more stable for varying temperatures. Indeed, MEMS-based amplitude SLMs are one of the most popular technology choices for many display applications, including projectors, so MEMS-based phase SLMs may also become increasingly important for holography applications. Unfortunately, the algorithms developed for high-precision LC-based phase SLMs \red{suffer from a degradation in image quality and fail to fully utilize time-multiplexing when used with the high framerate,} heavily quantized phase control that MEMS-based SLMs offer. For example, DLP’s phase SLM by Texas Instruments only offers up to 4~bits of precision or, similarly, 16~\red{unevenly distributed} discrete levels of phase control at frame rates of 1440~Hz \red{\cite{Bartlett:2019, ketchum2021diffraction}}. 

The focus of our work is to extend AI-driven CGH algorithms to operate with emerging fast but heavily quantized phase SLMs. This is a non-trivial task, because quantization is non-dif\-fer\-en\-ti\-a\-ble, so the standard machine learning toolset does not directly apply in these settings. Moreover, most of the degrees of freedom of a holographic display stem from their ability to create constructive and destructive interference, which can only be achieved instantaneously in time but not between time-multiplexed frames. 
It is thus not clear whether the partially-coherent holographic display mode enabled by the fast SLM speed is actually beneficial when combined with a limited precision of phase control or how it affects image quality. 
We propose an algorithmic CGH framework that robustly optimizes holograms in these mathematically challenging scenarios and explore the aforementioned tradeoff, demonstrating significant benefits in image quality and space--bandwidth utilization~\cite{Yoo:21} of higher-speed phase SLMs. Moreover, we develop a \red{learned propagation model} that is more flexible than previously proposed alternatives in allowing us to calibrate it using 3D multiplane supervision but leverage a variety of target content, including 2D images, 2.5D RGBD images, 3D focal stacks, and 4D light fields, for supervision during runtime. 

Specifically, our contributions include the following:
\begin{itemize}	
	\item a new variant of a camera-calibrated wave propagation model for holographic displays, which is flexible in enabling runtime supervision by 2D, 2.5D, 3D, or 4D content;
	\item a framework for robust CGH optimization with fast but heavily quantized phase-only SLMs;	
	\item experimental demonstration of improved image quality and better utilization of the SLM's space--bandwidth product enabled by our framework.
\end{itemize}
Source code for this paper is available at \MYhref{https://www.computationalimaging.org/publications/time-multiplexed-neural-holography/}{computationalimaging.org}.

%% file: sections/2_related.tex

Many aspects of holographic displays, including optics, SLMs, and algorithms, have advanced considerably over the last few years. Detailed discussions of many of these advancements can be found in the survey papers by Yaras~\shortcite{Yaras:2010}, Park~\shortcite{Park:2017}, and Chang et al.~\shortcite{Chang:20}. A recent roadmap article by Javidi et al.~\shortcite{Javidi:21} also outlines current and future research efforts of digital holography in non-display areas, including 3D imaging and microscopy.

Our work primarily focuses on advancing the algorithms driving holographic near-eye displays. In a nutshell, the CGH problem comprises several parts. First, the target content is specified in some format that needs to be converted to a complex-valued wavefield, such as point clouds~\cite{Gerchberg:1972,Fienup:1982,Shi:2017,Maimone:2017,shi2021towards}, polygons~\cite{Chen:2009,Matsushima:2009:polygon}, light rays~\cite{Wakunami:2013,Zhang:2011}, image layers~\cite{Chen:15,Zhang:2017,Chen:21}, or light fields~\cite{Benton:1983,Lucente:1995,Ziegler:2007,Kang:08,Padmanaban:2019}. Second, this wavefield needs to be encoded by a phase-only SLM, which can be achieved by fast, direct phase coding approaches~\cite{Hsueh:1978,Maimone:2017,Lee:1970} or slow, iterative solvers, such as classic Gerchberg--Saxton-type algorithms~\cite{Gerchberg:1972,Fienup:1982} or variants of stochastic gradient descent~\cite{Chakravarthula:2019,peng2020neural}. 

Yet, the simulated wave propagation models used by most of these CGH algorithms do not always model the physical optics faithfully, thereby degrading image quality.  Moreover, the computational complexity of these algorithms often prevents them from being practical in the power-constrained settings of a wearable computing system. Emerging artificial intelligence--driven CGH approaches have focused on addressing these limitations. For example, surrogate gradient methods that use a camera in the loop (CITL) for hologram optimization can significantly improve image quality~\cite{peng2020neural,choi2021optimizing,peng2021partiallycoherent}. Alternatively, differentiable wave propagation models can be learned to \red{calibrate for} the gap between simulated models and physical optics~\cite{peng2020neural,chakravarthula2020learned,choi2021neural3d,Kavakli:22}. Moreover, neural networks can be trained to enable real-time CGH algorithms~\cite{Horisaki:2018,peng2020neural,shi2021towards,Horisaki:21}.  


\begin{figure*}[t!]
	\centering
		\includegraphics[width=2.1\columnwidth]{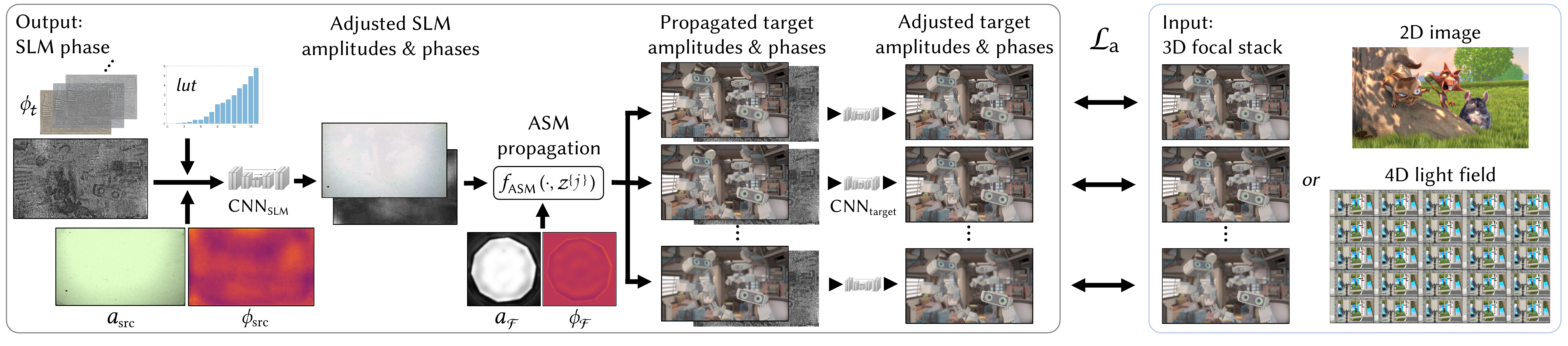}
		\caption{Illustration of our \red{calibrated} wave propagation model and 2D/3D/4D supervision strategy for the multiplexed, quantized hologram generation. The complex-valued field at the SLM is adjusted by several learnable terms (amplitude and phase at the SLM plane as well as look-up table for phase mapping) and then processed by a CNN. The resulting complex-valued wave field is propagated to all target planes using the ASM wave propagation operator with two extra learnable terms (amplitude and phase at the Fourier domain). The wave fields at each target plane are processed again by smaller CNNs. 
		The proposed framework applies to multiple input forms, including 2D, 2.5D, 3D, and 4D.}
		\label{fig:pipeline}
\end{figure*}
Note that our work is concurrently and independently developed from the very recent work by Lee et al.~\shortcite{lee2022high}. Although both works share some similarity in applying constrained gradient descent methods to optimize binary or heavily-quantized phase holograms, our framework outperforms the counterpart with the use of a learned propagation model for better image quality, the ability to effectively handle SLMs with varied bit depths and non-linear quantizations, and compatibility with a wide range of supervision sources.


%
%

%% file: sections/3_methods.tex

In Fresnel holography, a collimated coherent light beam illuminates an SLM with a source field $u_{\textrm{src}}$, and the light reflected in response reproduces a target intensity distribution.  To generate this hologram, a phase-only SLM imparts a spatially-varying delay $\phase$ on the phase of the field. After propagating a distance $z$ from the SLM, the resulting complex-valued field $u_z$ is given by the following image formation model:
\begin{align}
u_z \left( x, y, \lambda \right) &= \prop \left(\red{u_{\textrm{\tiny SLM}}\left( x, y, \lambda \right)}, z \right),  \nonumber \\
\red{u_{\textrm{\tiny SLM}}\left( x, y, \lambda \right)} &\red{= e^{i \quant \left( \phase \left( x, y, \lambda \right) \right)} u_{\textrm{src}} \left( x, y, \lambda \right), }
\label{eq:fresnel_holography}
\end{align}
where $\lambda$ is the wavelength of light, $x,y$ are the transverse coordinates, \red{and $u_{\textrm{\tiny SLM}}$ is the modulated field at the SLM.} The wave propagation operator $\prop$ models free-space propagation between two parallel planes separated by a distance $z$.
For notational convenience, we will omit the dependence on $x,y,\lambda$ and the source field $u_{\textrm{src}}$. The intensity pattern generated by this display at distance $z$ in front of the SLM when showing phase $\phase$ is therefore $\left| \prop \left( e^{i \quant \left( \phase \right)}, z \right) \right|^2$.

When \red{using low-bit SLMs for time-multiplexed holography}, the effect of quantization is not negligible.  To model a quantized phase-only SLM with $M \times N$ pixels, where every pixel offers phase control with limited precision, we define a quantization operator $\quant$:
\begin{equation} 
q: \mathbb{R}^{M \times N} \to \quantset^{M \times N}, \quad \phase \mapsto \quant(\phi) = \Pi_\quantset \left( \phase \right), 
\end{equation}
where $\Pi$ is the projection operator that maps the continuous phase value to the closest discrete phase in the feasible set $\quantset$ supported by the SLM.

Our framework approaches computer-generated holography with a differentiable camera-calibrated image formation model (Sec.~\ref{sec:model}), an optimization procedure designed for quantized SLMs (Sec.~\ref{sec:quantized}), and a family of loss functions supervised on either 2D, 2.5D, 3D, or 4D content to produce time-multiplexed holograms (Sec.~\ref{sec:supervision}). Figure~\ref{fig:pipeline} illustrates our model and optimization pipeline.

\subsection{Camera-calibrated Wave Propagation Model}
\label{sec:model}

Recent work on holographic displays has demonstrated that \red{the naive application of simulated wave propagation models, like the angular spectrum method (ASM)~\cite{Goodman:2014}, to holographic displays fails to account for the non-idealities of the} physical optical system, such as phase distortions of the SLM, optical aberrations, and the limited diffraction efficiency of the SLM~\cite{peng2020neural,chakravarthula2020learned,choi2021neural3d}. This discrepancy between simulated and physical image formation adversely affects image quality, but can be overcome by learning \red{to calibrate for} the physical optics using a differentiable, neural network--parameterized propagation model. 

Here, we propose a variant of the learned model recently proposed by Choi et al.~\shortcite{choi2021neural3d}:
\begin{align}
	 \propModel \! \left( \red{u_{\textrm{\tiny SLM}}}, z \right) \! & = \! \cnntarget \! \left( \propASM \! \left( \cnnslm \! \left( \slmAmpParam e^{i \slmPhaseParam} \red{u_{\textrm{\tiny SLM}}} \right), z \right) \right), \nonumber \\
	\propASM & \left( u,  z \right)  = \iint \Fourier \left( u \right) \cdot \transfer \left(f_x, f_y, \lambda, z \right) e^{ i 2 \pi (f_x x + f_y y)}d f_x d f_y \, , \nonumber \\
 \transfer &  \left(f_x, f_y, \lambda, z \right)  =  a_{\tiny \Fourier} \, e^{ i \left( \frac{2 \pi}{\lambda} z\sqrt{1 - \left( \lambda f_x \right)^2 - \left( \lambda f_y \right)^2} + \phase_{\tiny \Fourier} \right) } , 
\label{eq:model}
\end{align}
where $\cnnslm$ and $\cnntarget$ are convolutional neural networks that operate on the complex field at the SLM and target planes.  The target plane is a distance $z$ from the SLM. In addition, $\slmAmpParam$ and $\slmPhaseParam$ are learned to account for content-independent spatial variations in amplitude and phase of the incident source field at the SLM plane while $a_{\tiny \mathcal{F}}$ and $\phi_{\tiny \mathcal{F}}$ are added to the ASM propagation to learn spatial variations in amplitude and phase in the Fourier plane similarly to the learned complex convolutional kernel presented by~\citet{Kavakli:22}. 

Similar to Choi et al., we capture a training and a test set comprised of a large number of SLM phase patterns and corresponding amplitude images recorded at a set of distances $\left\{ j\right\}, j=1 \ldots J$ with our prototype holographic display. Using a standard stochastic gradient descent--type solver, we then fit the parameters of the CNNs, $\cnnslm$ and $\cnntarget$, as well as $\slmAmpParam, a_{\Fourier}, \slmPhaseParam, \phase_{\Fourier}$ to learn the \red{calibrated} wave propagation model. 
The model used in this framework builds upon the model from Choi et al. by using the terms $\slmAmpParam$, $\slmPhaseParam$, $\phi_{\tiny \mathcal{F}}$, and $a_{\tiny \mathcal{F}}$ to learn many of the content-independent non-idealities of the holographic system. The \red{source terms} can efficiently model the effects of non-ideal illumination at the SLM plane, and the Fourier plane terms can compactly account for the effects of non-ideal optical filtering. Together these terms enable the use of smaller convolutional neural networks to learn the content-dependent non-idealities, \red{such as the spatially varying pixel response at the SLM.}
Table~\ref{tab:ablation} quantitatively assesses the effect of these physically-inspired parameters by evaluating the performance of different \red{calibrated} wave propagation models on a captured dataset. All models are trained over 6 intensity planes, corresponding to 0.0~D, 0.5~D, 1.0~D, 1.5~D, 2.5~D, and 3.0~D in the physical space.  A $7^{\textrm{th}}$ plane at 2.0~D is set as the held-out plane for evaluation. In this table, we also ablate the performance of an additional $lut$ parameter to optionally learn the feasible set $\quantset$ of quantized values supported by the SLM. We observe that our model (bottom row) significantly reduces the number of parameters when compared to the original NH3D model, while still producing the highest PSNR metrics on the test set and the held-out plane.  \red{Notably, the lagging performance of the NH model, which is purely composed of physically-inspired terms, illustrates the substantial benefit of incorporating the flexibility of CNNs in a calibrated propagation model.} \red{Further details on our model architecture and training are included in Supplement S2.4}

\begin{table}
\footnotesize
\centering
\caption{\label{tab:ablation}Comparison of different \red{calibrated} wave propagation models. All models are trained on 6 of the 7 planes. PSNR is evaluated for training and test sets as well as for the 7$^{\textrm{th}}$ held-out plane. The number of parameters of each model is also reported. Training details are \red{listed} in Supplement S2.4.
}
\begin{tabular}{l|cccc}
\toprule
\rowcolor{gray!25}
 Models & Params. & Train & Test & Held-out \\
\midrule
NH~\cite{peng2020neural}                & 4.1M  & 26.7 & 27.1 & 26.3  \\
NH3D~\cite{choi2021neural3d}         & 68.5M & 34.4 & 32.4 & 31.9 \\
Our model, CNNs only								& 6.2M & 31.6 & 29.7 & 30.0 \\
  \; + $\slmAmpParam$                                                          & 7.2M & 35.3  & 35.4 & 32.3 \\
  \; + $\slmAmpParam$ + $\slmPhaseParam$                                 & 8.2M & 36.2 & 36.3 & 33.0 \\
  \; + $\slmAmpParam$ + $\slmPhaseParam$ + $\phi_{\tiny \mathcal{F}}$   &12.3M & 36.5 & 36.4 & 32.8 \\
  \; + $\slmAmpParam$ + $\slmPhaseParam$ + $\phi_{\tiny \mathcal{F}}$ + $lut$     & 12.3M & 36.4 & 36.4 & 32.8 \\
  \; + $\slmAmpParam$ + $\slmPhaseParam$ + $\phi_{\tiny \mathcal{F}}$ + $a_{\tiny \mathcal{F}}$ + $lut$   & 16.4M & \bf 36.7 & \bf 36.7 & 32.6 \\
\bottomrule
\end{tabular}
\end{table}

\subsection{Optimizing Phase Patterns for Quantized SLMs}
\label{sec:quantized}

Emerging MEMS-based phase SLMs are fast but offer only a limited precision for controlling phase. DLP’s phase SLM by Texas Instruments (TI)~\cite{Bartlett:2019}, for example, runs at a maximum framerate of 1440~Hz grayscale but only offers 4 bits, or 16 discrete phase levels, at each of the frames.
We therefore need to derive methods that allow us to optimize phase patterns for heavily quantized phase SLMs. The primary problem is that the quantization function $\quant$ is not differentiable. To this end, we discuss and evaluate several strategies for dealing with $\quant$ assuming some simple 2D loss function $\loss\left( s \cdot \Big| \propModel \left( e^{i \quant \left( \phase \right) }, 0 \right)  \Big|, \target  \right)$, where  \red{$\target$ is the desired 2D amplitude, and $s$ is a scale parameter that is optimized along with $\phase$.}  

\begin{figure}[tb]
	\centering
		\includegraphics[width=\columnwidth]{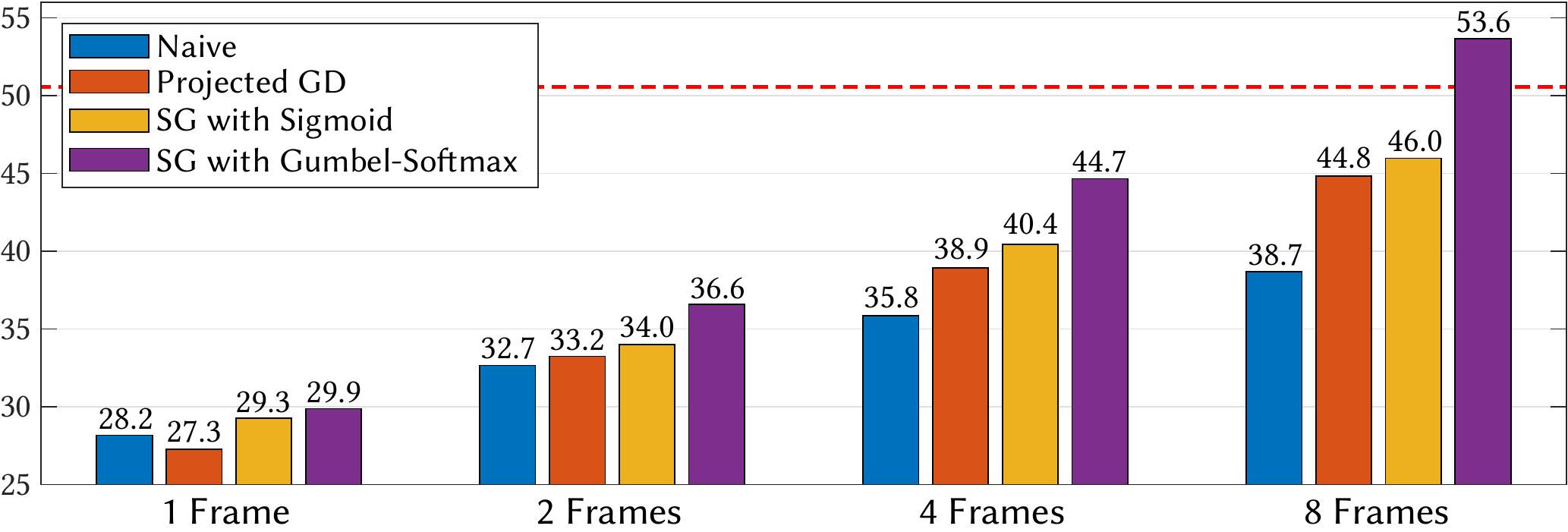}
		\caption{Evaluation of CGH algorithms for fast, heavily quantized phase SLMs. We show simulations of 4~bit phase quantization with varying numbers of time-multiplexed frames, showing the average PSNR over 14 example images. The projected gradient descent (GD) improves upon the naive method, which ignores quantization. Surrogate gradient (SG) methods replace the gradients of the non-differentiable quantization operator in the backpropagation pass using either a sigmoid or a Gumbel-Softmax (GS) function. The latter is found to outperform other approaches by a large margin, especially with faster SLMs. Remarkably, our framework using only 4 bit precision with 8 time-multiplexed frames even outperforms a conventional 8~bit phase SLM without time multiplexing (red dashed line).}
		\label{fig:comparison}
\end{figure}

The naive solution to dealing with $\quant$ is to simply ignore it. Specifically, the phase pattern $\phase$ can be optimized given a 2D target amplitude image $\target$ and quantized to the available precision after the optimization. This is \red{the approach typically adopted by state-of-the-art CGH algorithms that work well for liquid crystal--type phase SLMs, because these SLMs offer 8~bit or higher precision phase modulation.} TI's MEMS device \red{enables time multiplexing} but only offers 4 bits, which makes this approach impractical (see Fig.~\ref{fig:comparison}). Instead, the reference code supplied with the SLM implements a variant of projected gradient descent~\cite{boyd2004convex}, which projects the iteratively updated solution onto the feasible set of quantized values $\quantset$. 
This approach is equivalent to a gradient descent--type update scheme that applies $\quant$ after each iteration $k$ as:
\begin{align}
	\widehat{\phase}^{(k)} & \leftarrow \phase^{(k-1)} - \alpha \! \left( \frac{\partial \loss}{\partial \phase } \right)^T \!\! \loss \left( s \cdot \big| \propModel \left( e^{i \phase^{(k-1)}} \right) \big|, a_{\text{target}} \right) , \nonumber \\
	\phase^{(k)} & \leftarrow \Pi_\quantset \left( \widehat{\phase}^{(k)} \right) = \quant \left( \widehat{\phase}^{(k)} \right). 
	\label{eq:phaseupdate}
\end{align}

As an alternative solution to solving these types of problems, surrogate gradient methods are often used~\cite{bengio2013estimating,zenke2018superspike}. Here, the forward pass is computed using the correct quantization function $\quant$ but during the error backpropagation pass, the gradients of a differentiable proxy function $\widehat{\quant}$ are used. \red{This enables improved optimization of phase patterns through a quantization layer with the minimal overhead of computing the proxy gradients:}
\begin{equation}
	\phase^{(k)} \! \leftarrow \! \phase^{(k-1)} \! - \! \alpha \! \left( \frac{\partial \loss}{\partial \quant } \cdot \frac{\partial  \widehat{\quant} }{\partial \phase} \right)^T \!\!\!\! \loss \left( s \cdot \big| \propModel \left( e^{i \quant \left( \phase^{(k-1)} \right) } \right) \big|, a_{\text{target}} \right).
	\label{eq:phaseupdate_surrogate}
\end{equation}
Perhaps the most common choice for $\widehat{\quant}$ is a sigmoid function, whose slope can be gradually annealed during training~\cite{bengio2013estimating,zenke2018superspike,chung2016hierarchical}. 

We propose the use of a continuous relaxation of categorical variables using Gumbel-Softmax~\cite{jang2016categorical,maddison2016concrete} for optimizing heavily quantized phase values in CGH applications. This approach has several desirable properties. First, the Gumbel noise and categorical relaxation prevent the optimization from getting stuck in local minima, which is perhaps the primary benefit over other surrogate gradient methods. Second, annealing of the temperature parameter $\tau$ of the softmax as well as the shape of the score function are directly supported. Formally, this approach is written as:
%
\begin{align}
\widehat{\quant} \left (\phi \right ) & = \sum_{l=1}^L \quantset_l \cdot \gumbel_l \left( \score\left( \phase, \quantset \right) \right), \\
\gumbel_l \left( z \right) & =  \frac{ \text{exp} \left( \left( z_l + g_l \right)/\tau\right)}{\sum_{l=1}^L \text{exp} \left( \left( z_l + g_l \right)/\tau\right)},  \\
\score_l \left( \phi, \quantset \right) & = \sigma \left( w\cdot \delta \left( \phi , \quantset_l \right) \right)\left( 1 -  \sigma \left( w\cdot \delta \left( \phi , \quantset_l \right) \right) \right),
\end{align}
where $g_l \sim \text{Gumbel} \left( 0, 1 \right)$ is the Gumbel noise for all of the $l = 1, \ldots, L$ categories, i.e., quantized phase levels, $\sigma$ is a sigmoid function, $\delta$ is the signed angular difference, and $w$ is a scale factor (see Jang et al.~\shortcite{jang2016categorical} \red{and the supplement} for additional details).

\subsection{Runtime Supervision of Time-multiplexed Holograms}
\label{sec:supervision}

Fast MEMS-based phase SLMs can produce higher-quality holograms through \red{time multiplexing}, i.e., intensity averaging of multiple frames.  Given our camera-calibrated wave propagation model (Sec.~\ref{sec:model}), we optimize for time-multiplexed holograms using different target content at runtime.



\paragraph{2D Holography} In this case, we wish to synthesize a 2D intensity image at a distance $z$ in front of the phase SLM. The distance can be fixed or dynamically varied in software to enable a varifocal holographic display mode. For this purpose, we specify the loss:
\begin{equation}
	\loss_{\tiny \textrm{{2D}}} = \loss \left( s \sqrt{ \frac{1}{T} \sum_{t=1}^T \Big| \propModel \left( e^{i \quant \left( \phase^{(t)} \right) } , z \right)  \Big|^2 }, \target  \right) ,
\end{equation}
between the target amplitude image $\target$ and the simulated holographic image and solve for $\phase$. We can easily formulate a time-multiplexed variant of the CGH problem using this loss function by summing over $t = 1 \ldots T$ squared amplitudes, i.e., intensities, where $T$ refers to the total number of time-multiplexed frames that can be displayed throughout the exposure time of the human eye. The simplest example of the loss function $\loss$ is an $\ell_2$ loss although other loss functions, such as perceptually motivated image quality metrics, could be applied as well.

\paragraph{2.5D Holography using RGBD Input}

Using the multiplane loss function presented by~\citet{choi2021neural3d}, holograms can be synthesized to generate a 2D set of intensities at depths specified by a depth map. We refer the interested reader to Supplement S2.5 for the loss function and an additional discussion on utilizing \red{time multiplexing} to produce natural blur with 2.5D supervision. 

\paragraph{3D Multiplane Holography}

True 3D holography can be achieved by optimizing a single SLM phase pattern $\phase$ or a series of time-multiplexed patterns $\phase^{(t)}$ for the target amplitude of a focal stack $\textrm{fs}_{\tiny \textrm{target}}$. The corresponding loss function in our framework looks very similar to that of the 2D hologram above, although it is evaluated over the set of focal slices $\left\{ j \right\}$:
\begin{equation}
	\loss_{\tiny \textrm{{3D}}} = \loss \left( s \sqrt{ \frac{1}{T} \sum_{t=1}^T \Big| \propModel  \left( e^{i \quant \left( \phase^{(t)} \right) } , z^{ \left\{ j \right\} } \right)  \Big|^2 }, \textrm{fs}_{\tiny \textrm{target}}  \right).
\end{equation}
Effectively optimizing this focal stack loss using the full blur available within the diffraction angle of the SLM requires \red{time multiplexing} as illustrated in Supplement S2.6.

\paragraph{4D Light Field Holography}

Finally, we can also supervise our CGH framework using the amplitudes of a 4D target light field $\textrm{lf}_{\tiny \textrm{target}}$. For this purpose, a differentiable hologram-to-light field transform is required, which can be calculated using the Short-time Fourier transform (STFT)~\cite{Zhang:2009,Padmanaban:2019}:
\begin{equation}
	\loss_{\tiny \textrm{{4D}}} = \loss \left( s \sqrt{ \frac{1}{T} \sum_{t=1}^T \Big| \textrm{STFT} \left( \propModel \left( e^{i \quant \left( \phase^{(t)} \right) } , z \right) \right)  \Big|^2 }, \textrm{lf}_{\tiny \textrm{target}}  \right).
\end{equation}
By utilizing {time multiplexing}, our optimized holograms can uniquely reproduce a set of light field views that fully covers the SLM's space--bandwidth product as detailed in Supplement S2.7.

%% file: sections/4_results.tex
%
%
\red{To evaluate our novel algorithms, we use a benchtop 3D holographic display prototype. This prototype includes a FISBA RGBeam fiber-coupled module with red, green, and blue optically aligned laser diodes for illumination and a TI DLP6750Q1EVM phase SLM for high-speed quantized phase modulation. We capture the images produced by this prototype with a FLIR Grasshopper3 12.3 MP color USB3 sensor through a Canon EF 35mm lens with focus controlled by an Arduino microcontroller. Further details of the prototype are included in Supplement S1.}

\begin{figure}[t!]
	\centering
		\includegraphics[width=\columnwidth]{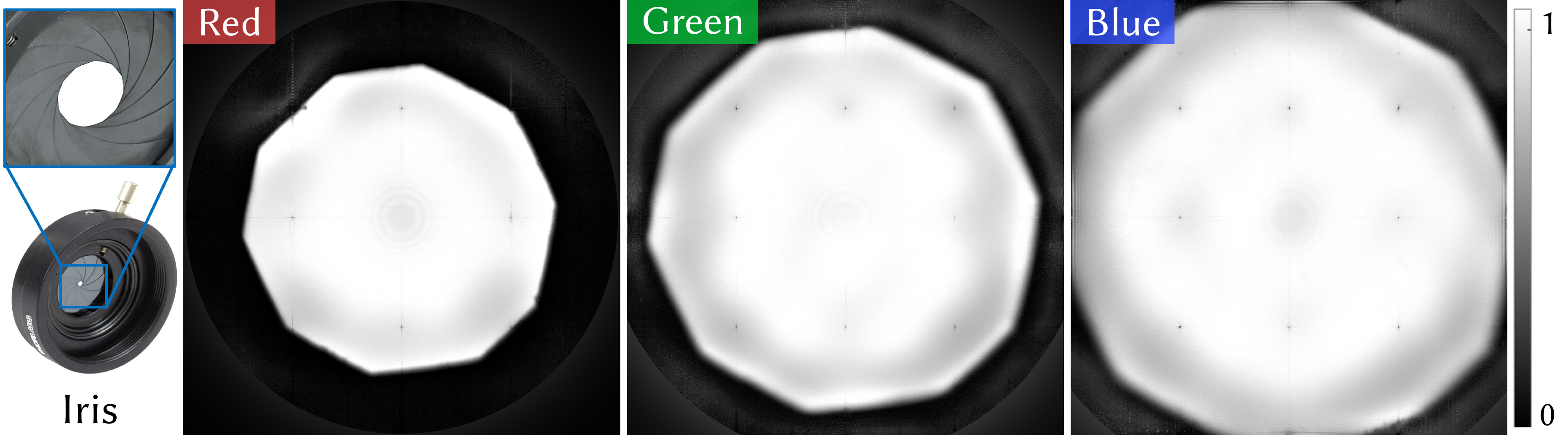}
		\caption{\red{Learned} optical filters for three channels, corresponding to the amplitude distribution on the Fourier plane $a_{\mathcal{F}}$ that is indicated in Sec.~3.1 and Table~1. On the left we show the photograph of the physical iris used in the system acting as the optical filter. Our model accurately learns the shape of the physical iris and, as expected, its diameter in the learned model varies accordingly to wavelength.}
		\label{fig:model_iris_vis}
\end{figure}

\begin{figure*}[t!]
	\centering
		\includegraphics[width=2.1\columnwidth]{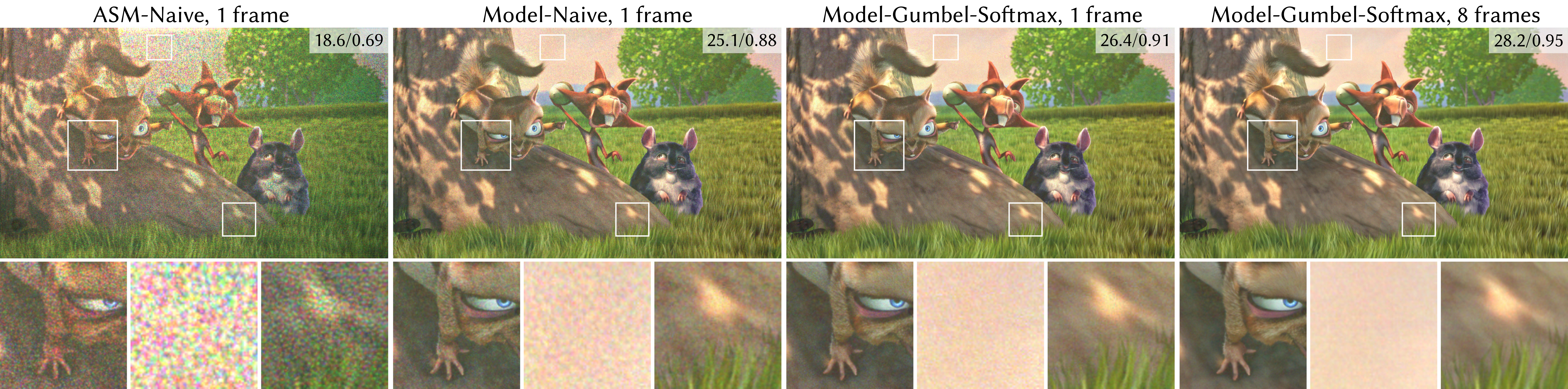}
		
		\caption{Comparison of 2D CGH algorithms using experimentally captured data. Here, we compare \red{SGD algorithms using the ASM w/ Naive (1st column), Model w/ Naive (2nd column), and Model w/ GS without time multiplexing (3rd column) and with 8 multiplexed frames (4th column). Our \red{calibrated} wave propagation model and Gumbel-Softmax quantization layer result in sharper images with higher contrast and less speckle than others under the same experimental conditions. Quantitative evaluations are included as PSNR/SSIM. }}
		\label{fig:results_2D}
\end{figure*}

\paragraph{Comparing CGH Algorithms}
We compare several CGH approaches for the task of optimizing phase patterns for a fast phase SLM with 4~bits, or 16 phase levels, in Fig.~\ref{fig:comparison}. The naive approach, which quantizes the phase after optimization performs poorly, as measured by the peak signal-to-noise ratio (PSNR). 
The projected gradient descent approach performs better and shows improvements with an increasing SLM speed. The surrogate gradient (SG) method used with the gradients of sigmoid and those of the Gumbel-Softmax are significantly better than other methods, with Gumbel-Softmax outperforming all other methods by a large margin, especially for higher-speed SLMs. This experiment represents the TI SLM with 4~bits and up to 480~Hz color, i.e., 8 multiplexed frames each running at 60~Hz so a total of 480~Hz. We evaluate other bit depths in the supplement and show similar trends. Finally, Gumbel-Softmax can be used as part of an SG method (Eq.~\ref{eq:phaseupdate_surrogate}) using only its gradients $\frac{\partial \widehat{\quant} }{\partial \phase}$ or it can be used to replace $\quant$ by $\widehat{\quant}$ also in the forward image formation. We found the former performs better in most settings, and therefore only report these results in the paper; see the supplement for evaluations of the latter approach.

\paragraph{Learning Physical Filters}
We visualize in Figure~\ref{fig:model_iris_vis} the performance of our learned model in accurately approximating the optical filter, which is an iris in the physical display system.
As expected, values outside the filters are all zeros. The shape of blade edges is robustly learned with our model and scales with wavelength as expected. The variance of diameter size also aligns with the variance of wavelength.
Refer to Figure~S7 in the supplement for visualization of the full model.

\paragraph{Assessing 2D Holography}
We present in Figure~\ref{fig:results_2D} experimental results of 2D holographic display assessing different CGH algorithms and different multiplexing schemes. 
\red{In this experiment, we compare SGD algorithms using the ASM with Naive quantization, our model with Naive quantization, and our model with Gumbel-Softmax (GS).}
We observe two insights. First, \red{the use of our calibrated} wave propagation model corrects for most artifacts present in the physical display. Second, applying the GS operation
leads to better performance in such heavily-quantized optimization problems.
Refer also to Figures~S8--9, as well as Tables~S1 and S2 in the supplementary document for both quantitative and qualitative assessments of other examples.

\begin{figure*}[t!]
	\centering
		\includegraphics[width=2.1\columnwidth]{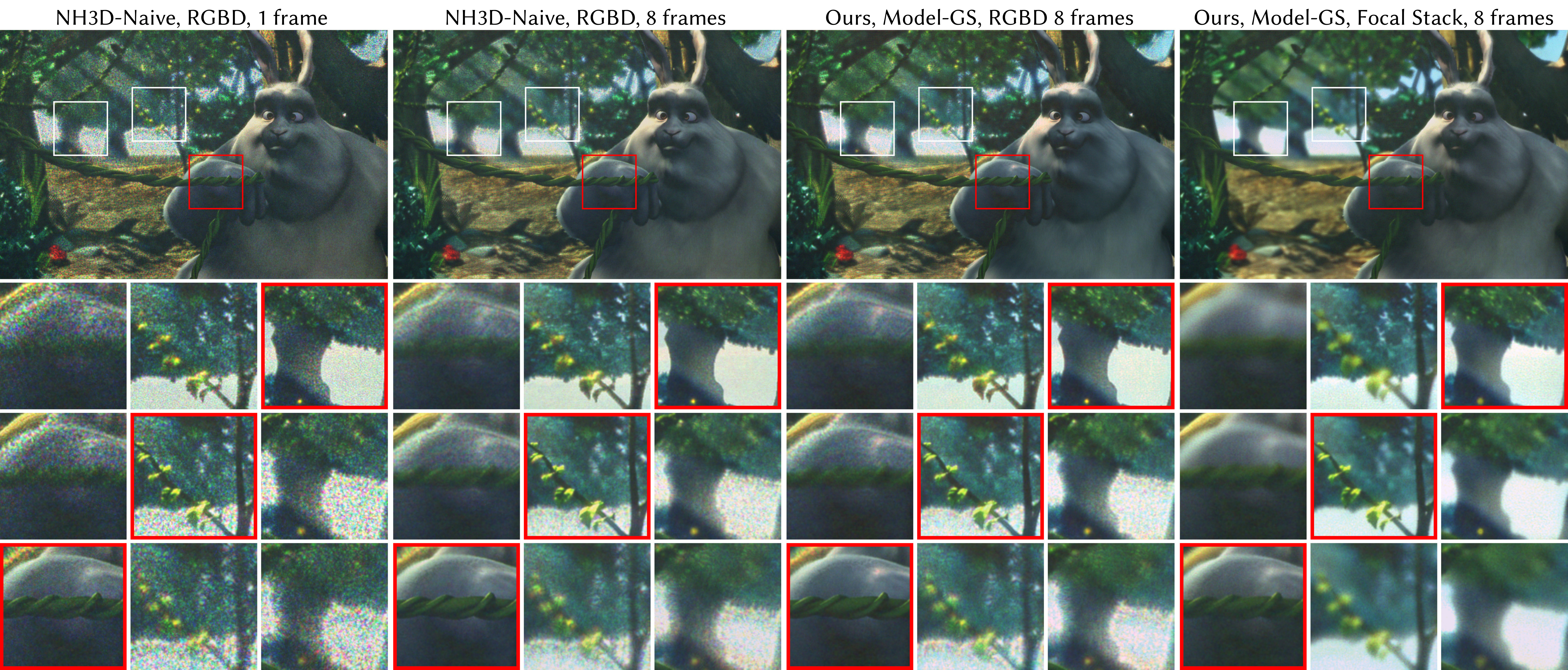}
		\caption{Comparison of 3D CGH algorithms using experimentally captured data. Here, we compare \red{SGD algorithms with the prior state-of-the-art NH3D model and Naive quantization} using RGBD input~\cite{choi2021neural3d} with 1 frame and 8 multiplexed frames, respectively, \red{our model with Gumbel-Softmax (GS), and our model with GS using focal stack supervision.} The corresponding PSNR metrics are 24.3~dB, 25.8~dB, and 26.7~dB with respect to the RGBD all-in-focus targets (left 3 columns), and 26.9~dB with respect to the focal stack (right column).
For close-ups, red squares indicate where the camera is focused at three distances (from top to bottom: far, intermediate, and near). }
		\label{fig:results_3D}
\end{figure*}

\paragraph{Assessing 3D Holography}
We present in Figure~\ref{fig:results_3D} experimental results of 3D holographic display assessing different CGH algorithms.
In this experiment, we compare \red{SGD algorithms with the prior state-of-the-art NH3D model and Naive quantization} using RGBD input~\cite{choi2021neural3d} with 1 frame and 8 multiplexed frames, respectively, \red{our model with Gumbel-Softmax (GS), and our model with GS using focal stack supervision.} PSNR metrics are provided in the caption. \red{Using only a single frame results in speckly in-focus content (shown with red squares in Figure~\ref{fig:results_3D}). Even with multiple frames, RGBD supervision produces speckle in the unconstrained out-of-focus regions. However, with our focal stack supervision and \red{time multiplexing}, we observe natural out-of-focus blur, while still preserving sharpness for the in-focus content.} For example, the branch at the intermediate depth is sharp, and the sky in the background is smooth. In the supplement, we show extensive evaluations and ablations of 3D multiplane CGH methods for more 3D scenes (Figures~S3--4 and~S10--16).

\paragraph{Assessing 4D Light Field Holography}

\begin{figure*}[t!]
	\centering
		\includegraphics[width=2.1\columnwidth]{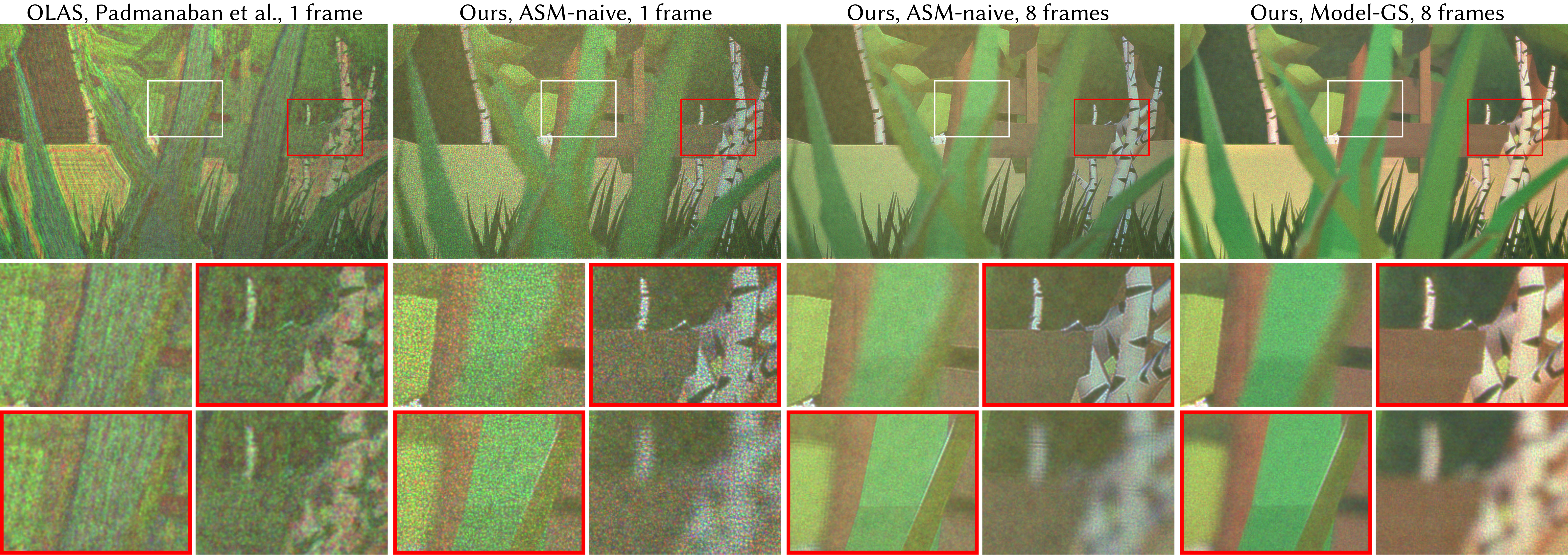}
		
		
		\caption{Comparison of 4D light field--supervised CGH algorithms using experimentally captured data. Here, we compare the OLAS algorithm~\cite{Padmanaban:2019} (1st column) without time multiplexing, and \red{three variants of our approach: ASM-Naive without time multiplexing (2nd column) and with 8 multiplexed frames (3rd column) and Model-GS with 8 multiplexed frames (4th column).} For close-ups, red squares indicate where the camera is focused at \red{two distances (top: far, bottom: near). Since OLAS deterministically computes a single phase pattern for a target light field, there would be no variation between time-multiplexed frames.}}
		\label{fig:results_4D}
\end{figure*}

We present in Figure~\ref{fig:results_4D} experimental results of 4D light field--supervised holographic display, assessing different CGH algorithms.
In this experiment, we compare the OLAS~\cite{Padmanaban:2019} algorithm, our approach using light field--supervision with the ASM and naive quantization (ASM-Naive), and our approach with the camera-calibrated wave propagation model and Gumbel-Softmax (Model-GS) to account for the low bit depth of the SLM. The OLAS algorithm requires light field and depth maps for each light field view as input and it does not support time multiplexing. Both variants of our method do not require depth maps and jointly optimize 8 time-multiplexed frames using SGD. For each example scene, we show close-ups of content at \red{two} distances \red{(far, near)}. We observe that our framework exhibits the best image quality for both in-focus (red squares) and out-of-focus regions (white squares). Refer also to Figures~S5 and~S17 in the supplementary document for additional simulation and experimental results.


%% file: sections/5_discussion.tex
In summary, we present a new framework for computer-generated holography. This framework includes a camera-calibrated wave propagation model that combines parts of the recently proposed model in a novel way to achieve a better performance with fewer model parameters. We explore surrogate gradient methods for optimizing the heavily quantized SLM patterns of emerging MEMS-based phase SLMs and show the Gumbel-Softmax algorithm to outperform other approaches. Our framework is flexible in supporting 2D, 2.5D, 3D, and 4D supervision at runtime and we show state-of-the-art results in all of these scenarios with our near-eye holographic display prototypes.

\paragraph{Limitations and Future Work}

Image quality could be further improved by increasing the precision and framerate of the employed phase SLMs and, importantly, by improving their diffraction efficiency. \red{In Figure S6 of our supplement, we explore the simulated image quality with varying levels of time multiplexing and bit depth, but analytically deriving this landscape remains an interesting direction for future work to explore.} Our algorithms do not run in real time, but require on the order of tens of seconds to a few minutes to compute a hologram. Neural networks could be employed to speed up the computation, as recently demonstrated by Horisaki et al.~\shortcite{Horisaki:2018}, Peng et al.~\shortcite{peng2020neural}, and Shi et al.~\shortcite{shi2021towards}. Due to their limited space--bandwidth product, holographic near-eye displays only provide a limited eye box, which could be addressed by dynamically steering it using eye tracking~\cite{Jang:2017}. The depth of field of 3D-supervised holograms in AR scenarios should match that of the user's eye, which requires tracking their pupil diameter. Finally, we demonstrated our results on benchtop prototype displays, which will have to be miniaturized into the impressive device form factors presented by Maimone et al.~\shortcite{Maimone:2017} and Wang and Maimone~\shortcite{Maimone:2020}.

\paragraph{Conclusion}

The algorithmic advances presented in this work help make holographic near-eye displays a practical technology for next-generation VR/AR systems.

%% file: sections/suppl_setup.tex

\begin{figure}[t!]
	\centering
		\includegraphics[width=\columnwidth]{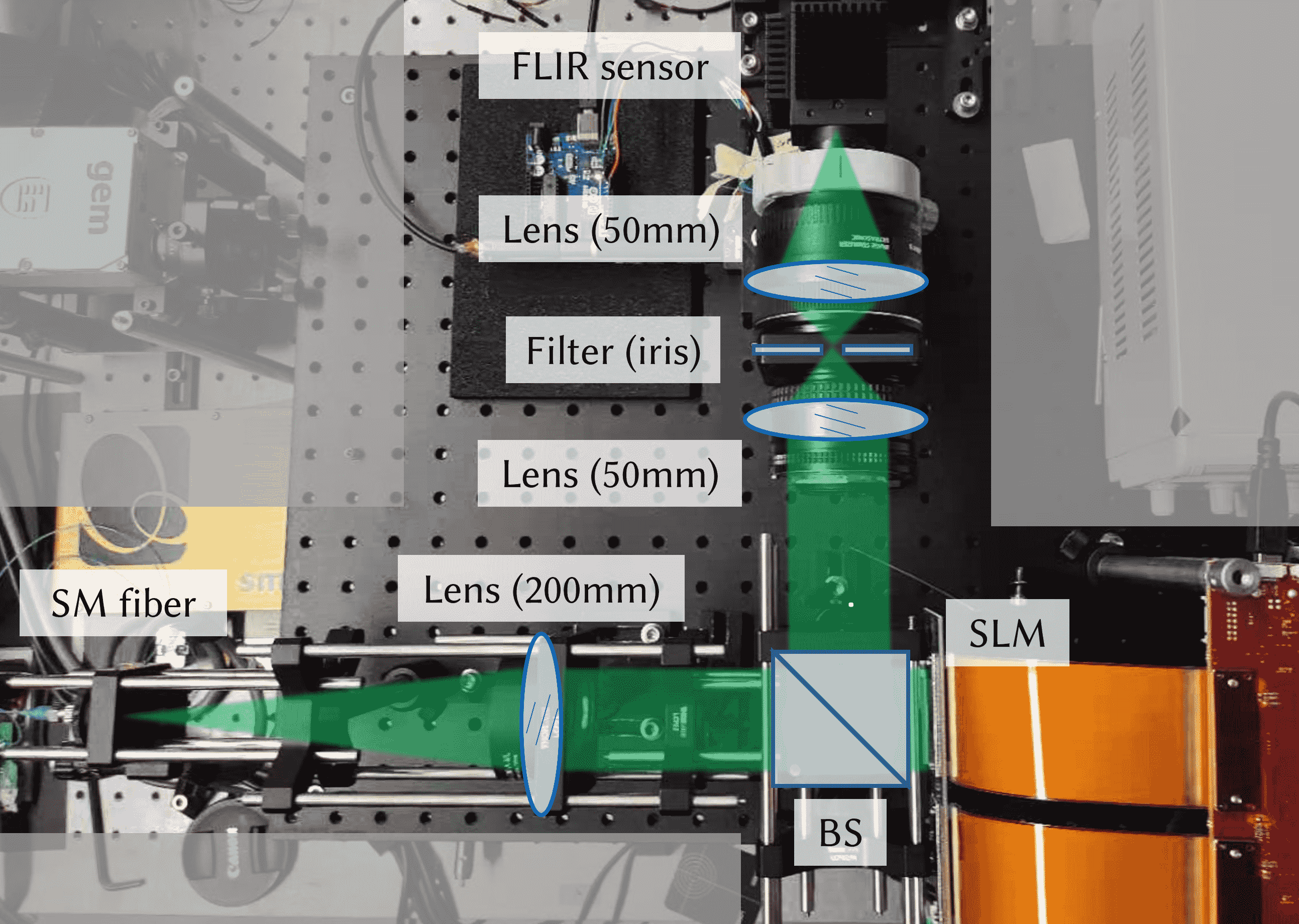}
		\caption{Schematic and prototype photograph of our holographic display.}
		\label{fig:prototype}
\end{figure}

In this section, we describe the hardware implementation of our benchtop 3D holographic display prototype.
Figure~\ref{fig:prototype} shows the system schematic and photograph of our implementation,
including a display and a capture unit, that are connected under a closed-loop framework. Specifically, the SLM is TI DLP6750Q1EVM 
with a resolution of 1,280 $\times$ 800, a pixel pitch of 10.8~$\mu$m, and a bit depth of 4~bits per pixel. 
The laser is a FISBA RGBeam fiber-coupled module with three optically aligned laser diodes with a maximum output power of 50~mW. 
The measured wavelengths are 636.4, 517.7, and 440.8~nm. In our implementation, color images are captured as separate exposures 
for each wavelength and then cast in post-processing.

Other components including the collimating lenses, the relay imaging lenses, the filtering iris, 
and the beam splitter (Thorlabs BS016) are shown in Figure~\ref{fig:prototype}.
All images are captured with a FLIR Grasshopper3 12.3 MP color USB3 sensor through a Canon EF 50mm lens. The Canon lens and sensor 
are synchronized in hardware via Arduino (Uno SMD) controller to enable programmable varifocal display and acquisition. The capture unit is
assembled on a motorized translation stage to enable the acquisition capability from different horizontal viewpoints. In such a way, we are
able to acquire holographic images to both form the training dataset and showcase diverse 3D cues.

In the calibration step, we use a similar procedure to that described in the relevant work~\cite{peng2020neural} and apply a planar homography from the field of 
computer vision to accurately register the captured images to the ground-truth images. Our implementation uses a target binary pattern
consisting of 18 $\times$ 11 white dots with the interval between the centers of neighboring two dots set 70 pixels. Accordingly, 
the region of interest has a resolution of 1,190 $\times$ 700 pixels. 




%% file: sections/suppl_alg.tex
\subsection{From a variant of projected gradient descent to surrogate gradient with unit Jacobian} 

Here, we derive the relationship between the variant of projected gradient descent in the manuscript and the gradient descent with the surrogate gradient of unit matrix Jacobian.

1) Based on the projected gradient descent rule described in the main paper (Eq.~4), we consider the projection step of the last iteration together with the phase update step of the current iteration as one iteration:
\begin{align}
	\phase^{(k-1)} & \leftarrow \Pi_\quantset \left( \widehat{\phase}^{(k-1)} \right) = \quant \left( \widehat{\phase}^{(k-1)} \right). \nonumber \\
	\widehat{\phase}^{(k)} & \leftarrow \phase^{(k-1)} - \alpha \! \left( \frac{\partial \loss}{\partial \phase } \right)^T \!\! \loss \left( s \big| \propCNN \left( e^{i \phase^{(k-1)}} \right) \big|, a_{\text{target}} \right). \nonumber
\end{align}

2) Then, we substitute the first row into the second row:
\begin{align}
	\widehat{\phase}^{(k)} & \leftarrow \quant\left( \widehat{\phase}^{(k-1)} \right)- \alpha \! \left( \frac{\partial \loss}{\partial \quant } \right)^T \!\! \loss \left( s \big| \propCNN \left( e^{i \quant \left( \widehat{\phase}^{(k-1)} \right)} \right) \big|, a_{\text{target}} \right). \nonumber
\end{align}

Note that all of these variants (including the one in the manuscript) generally fail to work well for holographic phase retrieval, because the projection taken every step vanishes the phase update with the gradient term. 

3) Thus, we relax this hard constraint by leaving out the projection in the first term; this variant of projected gradient descent applying the projection only in the second term does not suffer from being stuck. This is also a special case of surrogate gradient where we use the surrogate gradient of unit Jacobian $\frac{\partial \widehat{\quant}}{\partial \phase } = I$. This means the gradient with respect to the quantized phase $q$ is simply passed to the gradient with respect to the continuous phase $\phase$. We use the following algorithm as the representative of the projected gradient descent:

\begin{align}
	\widehat{\phase}^{(k)} & \leftarrow \widehat{\phase}^{(k-1)} - \alpha \! \left( \frac{\partial \loss}{\partial \quant } \right)^T \!\! \loss \left( s \big| \propCNN \left( e^{i \quant \left( \widehat{\phase}^{(k-1)} \right)} \right) \big|, a_{\text{target}} \right)  \\
	& = \widehat{\phase}^{(k-1)} - \alpha \! \left( \frac{\partial \loss}{\partial \quant } \cdot \frac{\partial \widehat{\quant}}{\partial \phase }  \right)^T \!\! \loss \left( s \big| \propCNN \left( e^{i \quant \left( \widehat{\phase}^{(k-1)} \right)} \right) \big|, a_{\text{target}} \right)  .
	\label{eq:projgd_to_sg}
\end{align}

Note that Eq.~\ref{eq:projgd_to_sg} is identical to Eq.~5 in the main paper. The recent paper by Lee et al.~\shortcite{lee2022high} uses hard-sigmoid as a surrogate gradient which has unit gradient within the valid range, and we note that this falls into the category of variants of gradients we describe in this section and our Gumbel-Softmax based approach outperforms it with a large margin as shown in Fig.~3.

\begin{figure}[t!]
\centering
\includegraphics[width=\columnwidth]{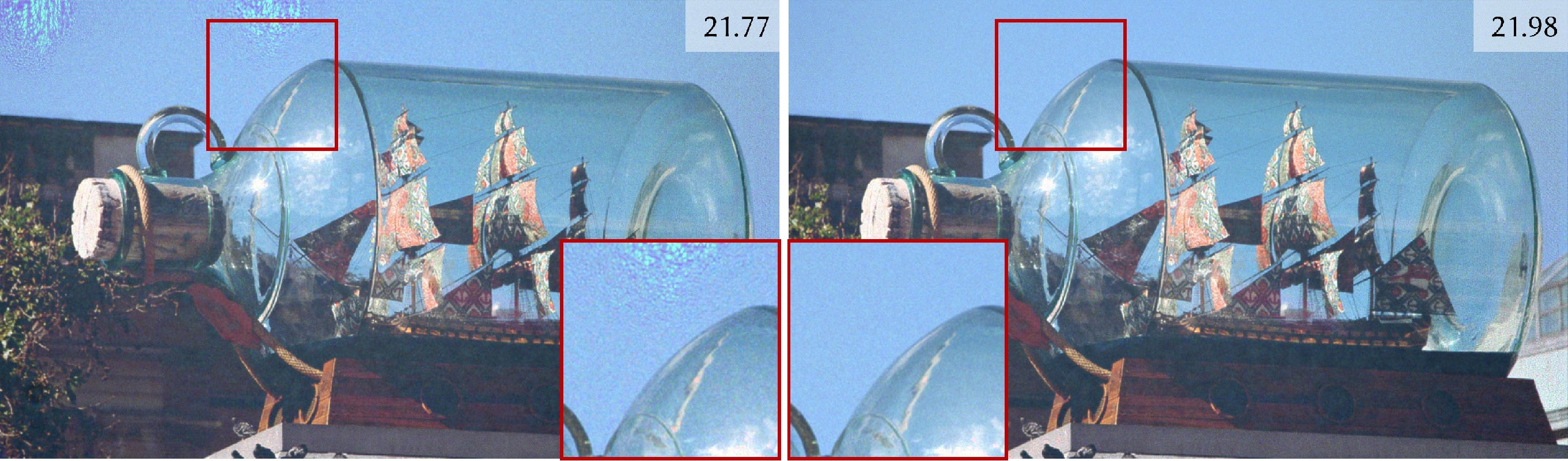}
\caption{Experimentally captured results with the camera-in-the-loop calibration using the naive (left) and the surrogate gradient (right) quantization. PSNR metrics are indicated.}
\label{fig:citl_sg_gs_naive}
\end{figure}

\subsection{Camera-in-the-loop with highly quantized SLMs}

Here we describe a more accurate camera-in-the-loop procedure for highly quantized SLMs. The camera-in-the-loop procedure proposed by Peng et al.~\shortcite{peng2020neural} approximates the gradient of the physical forward model $f$ to that of simulated model $\widehat{f}$:

\begin{align}
	\phase^{(k)} & \leftarrow \phase^{(k-1)} - \alpha \! \left( \frac{\partial \loss}{\partial \phase } \right)^T  \nonumber \\
	& \simeq  \phase^{(k-1)} - \alpha \! \left( \frac{\partial \loss}{\partial f } \cdot \frac{\partial \widehat{f} }{\partial \phase }  \right)^T   .
\end{align}

However, note that we always have to quantize the phase before displaying it on the SLM. Thus, technically, $f$ should reads as $f\left(\quant\left(\phi\right)\right)$. Again, while we do not have access to the gradient of the quantization function $\quant$, we can approximate it with the surrogate gradient $\frac{\partial \widehat{q}}{\partial \phase} $:

\begin{align}
	\phase^{(k)} & \leftarrow \phase^{(k-1)} - \alpha \! \left( \frac{\partial \loss}{\partial f\left(q\right) } \cdot \frac{\partial \widehat{f}\left(\quant\right) }{\partial \phase }  \right)^T   \nonumber \\
& = \phase^{(k-1)} - \alpha \! \left( \frac{\partial \loss}{\partial f } \cdot  \frac{\partial \widehat{f} }{\partial \quant } \cdot  \frac{\partial \quant }{\partial \phase }  \right)^T  \nonumber \\
& \simeq \phase^{(k-1)} - \alpha \! \left( \frac{\partial \loss}{\partial f } \cdot  \frac{\partial \widehat{f} }{\partial \quant } \cdot  \frac{\partial \widehat{\quant} }{\partial \phase }  \right)^T .
\end{align}

In Fig.~\ref{fig:citl_sg_gs_naive}, we compare two update rules Eq.~(3) and Eq.~(4). We see that with the approximation with the surrogate gradient, image quality is noticeably improved.

\subsection{Setting parameters for quantized phase optimization}
In this subsection we describe parameters used in quantized phase optimization. We run 2,000 iterations with early stopping with a learning rate of 0.01 for 1 frame and that of 0.02 for 8 frames. We note that the surrogate gradients, if used with the Sigmoid or functions with clipping, require a higher learning rate to avoid getting stuck in local minima which leads to poor performance. During optimization, gradually annealing the slope helps the optimization by better approximating the step function gradually while allowing exploration of a large parameter space with a lower slope at the beginning. The Sigmoid function can be annealed with a parameter $s$ multiplied with the input $x$, so we refer to the Sigmoid function as $\sigma \left( s \cdot x \right)$. The Gumbel-Softmax can be annealed with three parameters including the temperature parameter of Softmax $\tau$, the width parameter $w$ which corresponds to the interval of discrete levels, and a scale multiplied to the score function in Eq.~8 of the manuscript. In the experiments, we tuned $w$ considering the number of phase levels (interval between neighbour phase displacements) and the scale multiplied to the score function was increased from 300 to 1,000 during optimization. We used an annealing schedule of $\tau = \tau_0 \cdot e^{-c \cdot (t/t_\textrm{max}) }$ at iteration $t$ with $c \sim \textrm{ln}2$ and $\tau_0 \sim 4$.

\subsection{Model architecture and training details}
The two convolutional neural networks in our model are based on the U-net architecture as in Choi et al.~\shortcite{choi2021neural3d}. We made a slight modification on the CNN archtectures such that each CNN has 5 layers and 4 input channels of amplitude, phase, real, and imaginary values of the input field. The output of $\cnnslm$ is two channels that are used as real and imaginary values of the adjusted SLM field. The output of $\cnntarget$ is 1 channel that is used as a corrected amplitude. As stated in Table~1 of the main paper, the model is trained over 6 intensity planes, corresponding to 0.0~D, 0.5~D, 1.0~D, 1.5~D, 2.5~D, and 3.0~D in the physical space. The propagation distances from the SLM are 7.9, 8.1, 8.25, 8.4, 8.6, 8.8, 9.1 cm and the held-out plane is set to 8.6 cm. We use a batch size of 2, and a learning rate of $4e^{-4}$. We note that the variety of phasemaps are important for model training. For example, we note that a dataset mainly generated using the SGD algorithm usually consists of holographic images that have very narrow angular spectrum. Thus, we generate the dataset with the STFT-based regularizer we present in Eq.~S7. In addition, we generate phasemaps with a set of random parameters, including learning rates, initial phase distribution, and propagation distances. We generate 3,000 phases for each channel and capture the intensity at 7 target planes. Other than the held-out plane, the dataset is divided into training, validation, and test sets with a ratio of 8:1:1. The training takes around 24 hours to converge. To parameterize and train a look up table for phase mapping, the phase maps are first one-hot encoded, multiplied with the parameterized lookup table, and then summed up per pixel before passing through the full forward model pipeline.

\subsection{Natural defocus blur with 2.5D supervision on Quantized SLMs}
\begin{figure*}[t!]
	\centering
		\includegraphics[width=2.1\columnwidth]{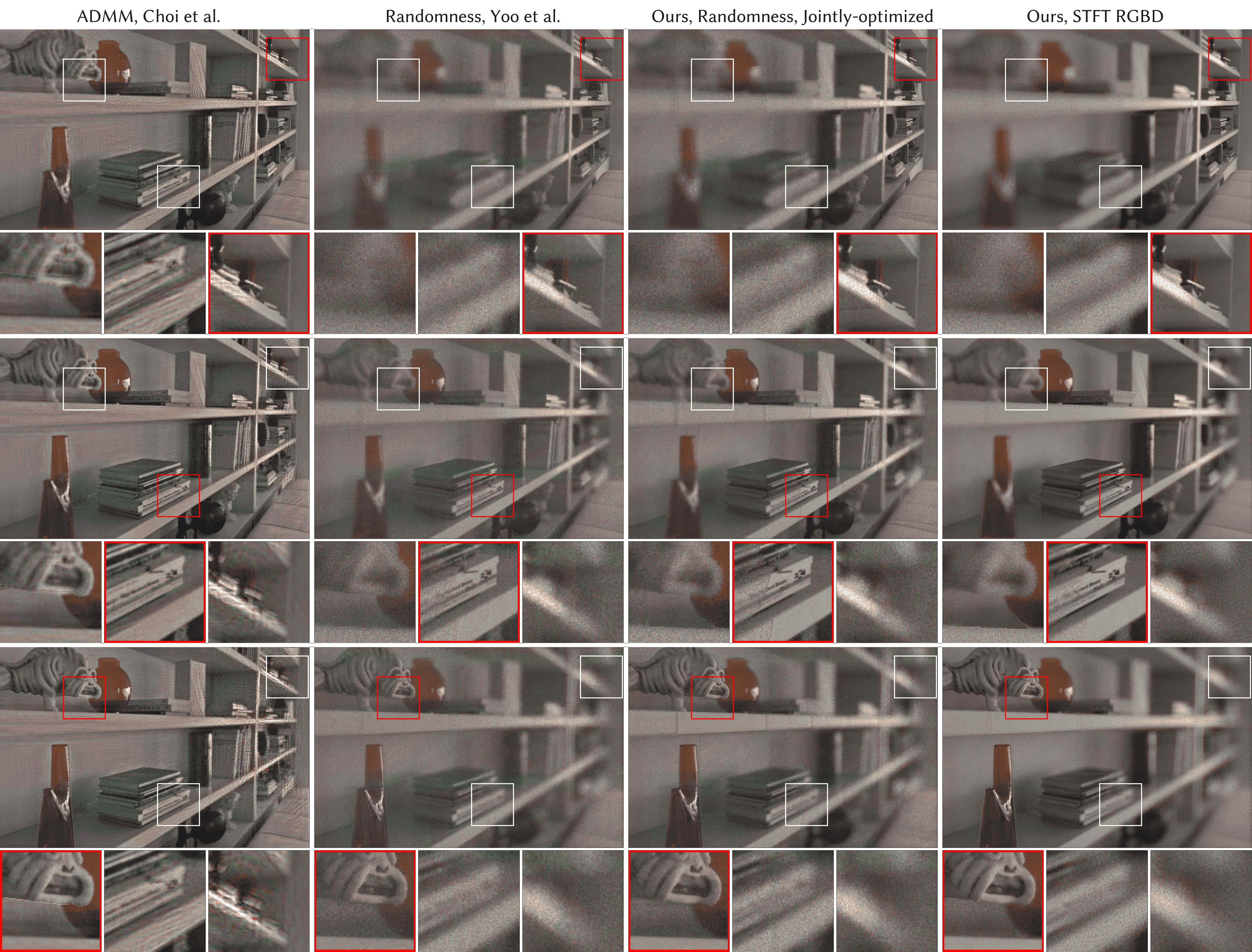}
		\caption{Simulated evaluation of different RGBD supervised techniques for generating smooth defocus blur on ideal quantized SLMs. From left to right: Model ADMM from Choi et al.~\cite{choi2021neural3d}, Randomness Prior from Yoo et al.~\cite{Yoo:21}, our Randomness Prior Jointly-optimized implementation, and our STFT RGBD implementation. Focused regions are highlighted with red boxes, that from top to bottom, indicate far, center, and near distances. The ADMM  phase smoothness technique from~\citet{choi2021neural3d} produces smooth but very small blur and has strong artifacts because it cannot be adapted to quantization effectively. The phase randomness technique with individually optimized frames produces more substantial blur but has poor in-focus image quality because it uses individually optimized frames. The phase randomness technique with jointly optimized frames improves on this in-focus image quality, and both of these phase randomness techniques suffer from out-of-focus artifacts on the quantized SLMs. Adding the STFT-based loss function greatly improves the out-of-focus blur and has high image quality.}
		\label{fig:rgbd_blur}
\end{figure*}

The 2.5D supervision results in our paper are generated using the multiplane loss function presented by~\citet{choi2021neural3d}. For this approach, the depth map $D$ from an RGBD input is first decomposed into a set of binary masks $m^{\{j\}}$ corresponding to a set of distances $z^{\{j\}}$ from the SLM using closest distance matching,
\begin{equation}
	m^{(j)}(x,y)=\begin{cases} 1, \quad \text{if } |z^{(j)}- D(x,y)| <  |z^{(k)}- D(x,y)|, \forall k\neq j, \\ 0, \quad \text{otherwise.}\end{cases}
\end{equation}
These binary masks are then used to constrain a multiplane loss that pushes the wavefront to reconstruct the desired RGB amplitude, $\target$, at the corresponding in-focus distances from the SLM
\begin{align}
	\loss_{\tiny \textrm{{2.5D}}} = \frac{1}{J} \sum_{j=1}^J \loss &\Bigg( m^{(j)} \circ  s \sqrt{ \frac{1}{T} \sum_{t=1}^T \Big| \propModel  \left( e^{i \quant \left( \phase^{(t)} \right) } , z^{ \left( j \right) } \right)  \Big|^2 }, \nonumber\\
& m^{(j)} \circ \target  \Bigg) ,
\end{align}
where $\circ$ is element-wise multiplication. One challenge with this approach is that it leaves the out-of-focus parts of the displayed intensities volume unconstrained, but this can be addressed using additional smooth phase regularization strategies as discussed further by~\citet{choi2021neural3d}. However, the ADMM technique for smooth phase proposed by this work can only produce very slight blur, and cannot effectively be adapted to quantization. 

Alternatively, with time multiplexing, some prior works including~\cite{Yoo:21} have proposed phase randomness approaches that can produce a much shallower depth of field. These techniques aim to randomly send light in differerent directions from each scene point. Over many frames, this results in scene points that diffusely send light in all directions. Unfortunately, this technique struggles with producing good image quality in the presence of quantization because it independently optimizes frames. Quantization also adds artifacts to the out-of-focus blur with this technique. To overcome these quantization artifacts and reduce the number of frames needed for smooth out-of-focus blur, an additional STFT-based loss can be applied to the in-focus content.
\begin{equation}
	\loss_{\tiny \textrm{STFT}} = \frac{1}{J} \sum_{j=1}^J \overline{\sigma_\theta^2} \left(m^{(j)} \circ s \sqrt{ \frac{1}{T} \sum_{t=1}^T \Big| \textrm{STFT} \left( \propModel \left( e^{i \quant \left( \phase^{(t)} \right) } , z \right) \right)  \Big|^2 } \right) ,
\end{equation}
where $\overline{\sigma_\theta^2}$ is the variance of the STFT over angles averaged over the spatial locations across the wavefront. This loss pushes the output of the holographic display to emit light evenly in all directions from the in-focus points. This mimics the diffuse behavior of most natural coherent scenes. As demonstrated in Fig.~\ref{fig:rgbd_blur}, this enables natural blur with 2.5D supervision on quantized SLMs.

\subsection{Time-multiplexing for 3D supervision}
\begin{figure*}[t!]
	\centering
		\includegraphics[width=1.7\columnwidth]{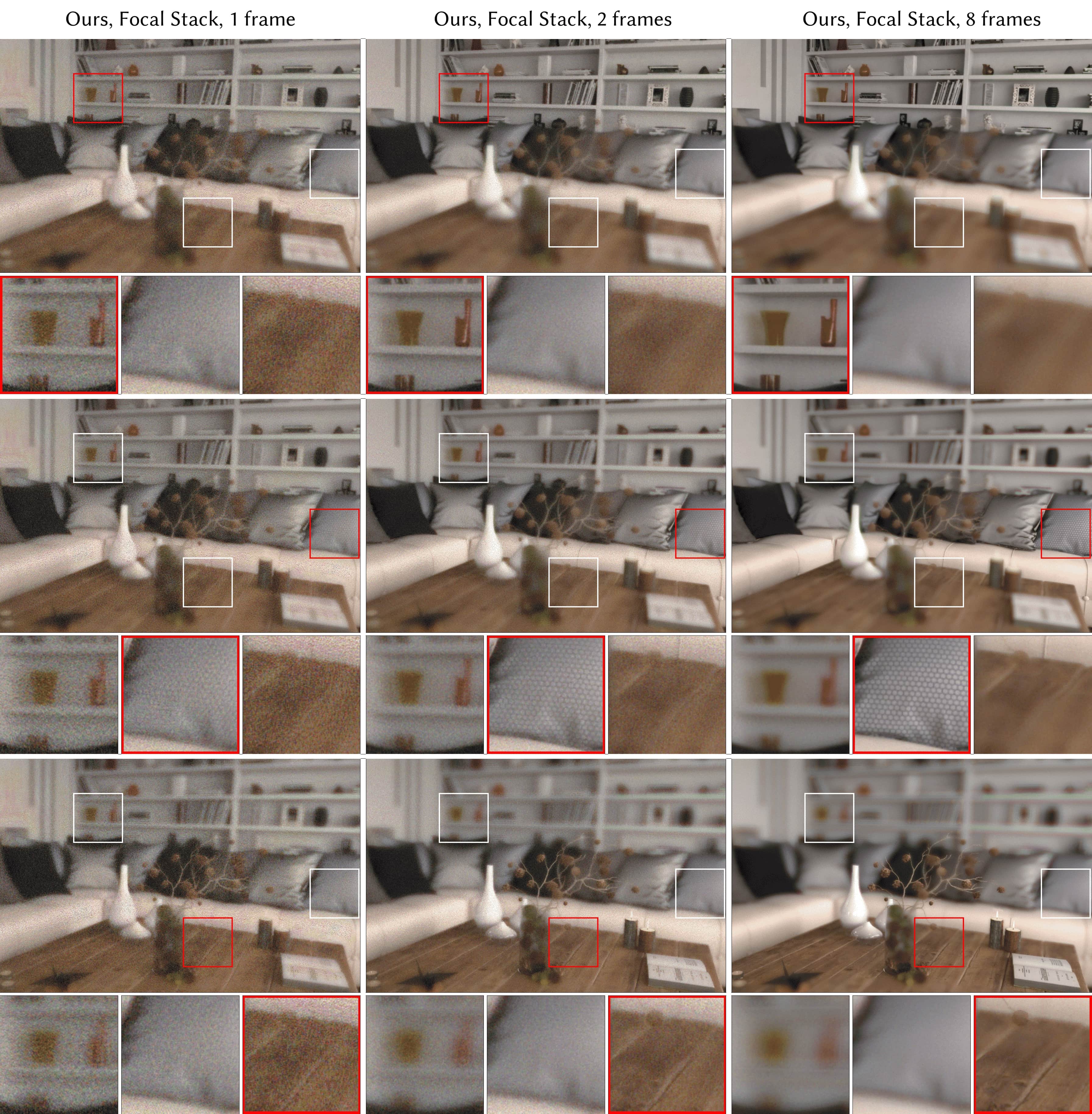}
		\caption{Simulated evaluation of focal stack supervision with 1, 2, and 8 frames from left to right on ideal continuous SLMs. Focused regions are highlighted with red boxes, that from top to bottom, indicate far, center, and near distances. Even without quantization on an ideal SLM, the supervision with only 1 or 2 frames is overconstrained by the focal stack and cannot fully reproduce the desired natural defocus blur.}
		\label{fig:focalmultiplex}
\end{figure*}
Our 3D focal stack supervision technique enables very high image quality with natural defocus effects. This technique relies on the time multiplexing in order to reproduce the defocus effects as illustrated in Fig.~\ref{fig:focalmultiplex}. Some prior works such as~\citet{shi2021towards} have used similar focal stack supervision with a single frame but that is only possible with much less blur. This blur is produced by a low frequency coherent wavefront and cannot match the natural blur produced by a scene sending light in all directions.

\subsection{Time-multiplexing for 4D supervision}
\begin{figure*}[t!]
	\centering
		\includegraphics[width=2.1\columnwidth]{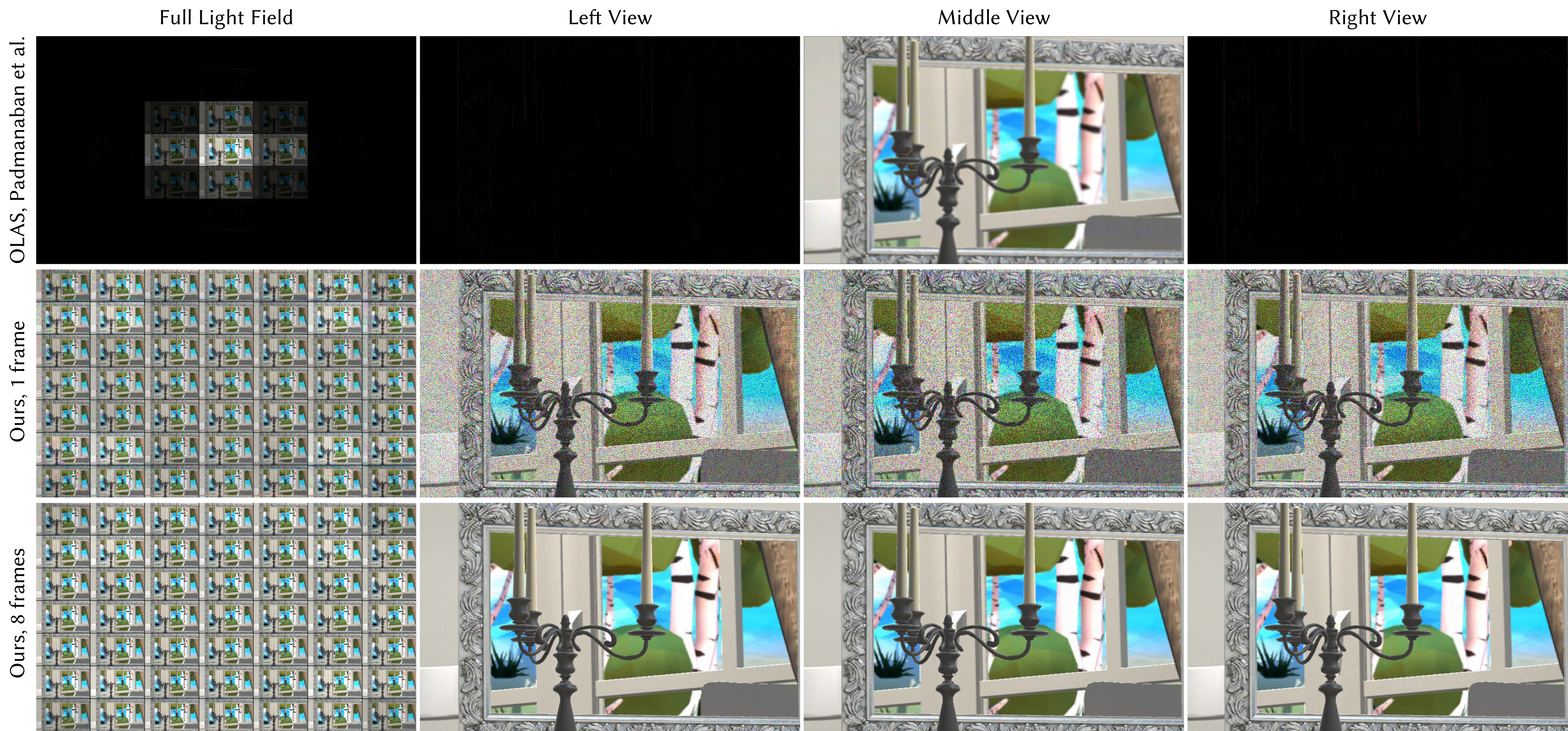} 
		\vspace*{6pt}
		\includegraphics[width=2.1\columnwidth]{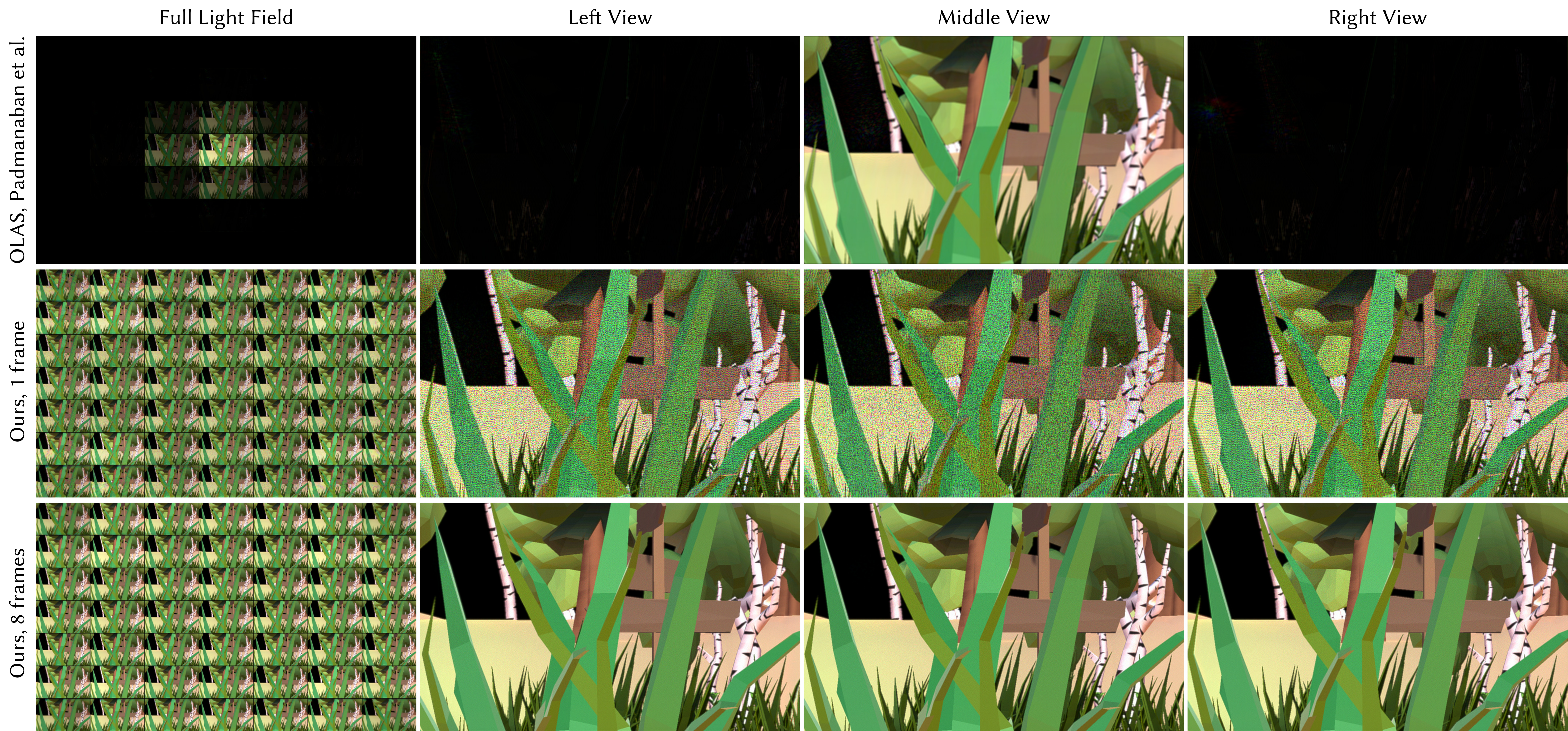}
		\caption{Simulated evaluation of different light field to hologram techniques on ideal continuous SLMs. In the first column we have the full set of reproduced light field views. From the second column to the fourth column, we present the comparison of selected views. (Top) OLAS by Padmanaban et al.~\shortcite{Padmanaban:2019} which does not account for the interference of rays has heavily amplified smooth content in the central view and heavily attenuated rays in other views. (Middle) Even without quantization on an ideal SLM, our proposed light field supervision technique with a single frame better covers the light field views but lacks the degrees of freedom to reproduce all the rays across all the light field views. (Bottom) Our proposed technique jointly over 8 frames has the degrees of freedom to fully reproduce the light field views.}
		\label{fig:lfmultiplex}
\end{figure*}

Our approach can uniquely use a holographic display to reproduce a full set of light field views. Prior holographic stereogram works did not account for how interference attenuates and amplifies different rays after converting a light field into a hologram. With the recently proposed overlap-add stereogram (OLAS) method~\cite{Padmanaban:2019}, this interference results in most rays outside of the central light field view being heavily attenuated by destructive interference. Additionally, smooth content in the central view gets amplified by constructive interference. Along with modeling this interference, time multiplexing is needed to accurately reproduce a set of light field views that fully cover the diffraction angle of the SLM. Without time multiplexing, a single coherent wavefront produced with phase modulation of the SLM's resolution will not have the degrees of freedom to produce arbitrary light field views that could naturally occur. The phenomena discussed here is illustrated in Fig.~\ref{fig:lfmultiplex}.


%% file: sections/suppl_results.tex

In this section, we present extra simulated and experimental results of our 3D holographic display prototype.

\begin{figure*}[t!]
	\centering
		\includegraphics[width=2\columnwidth]{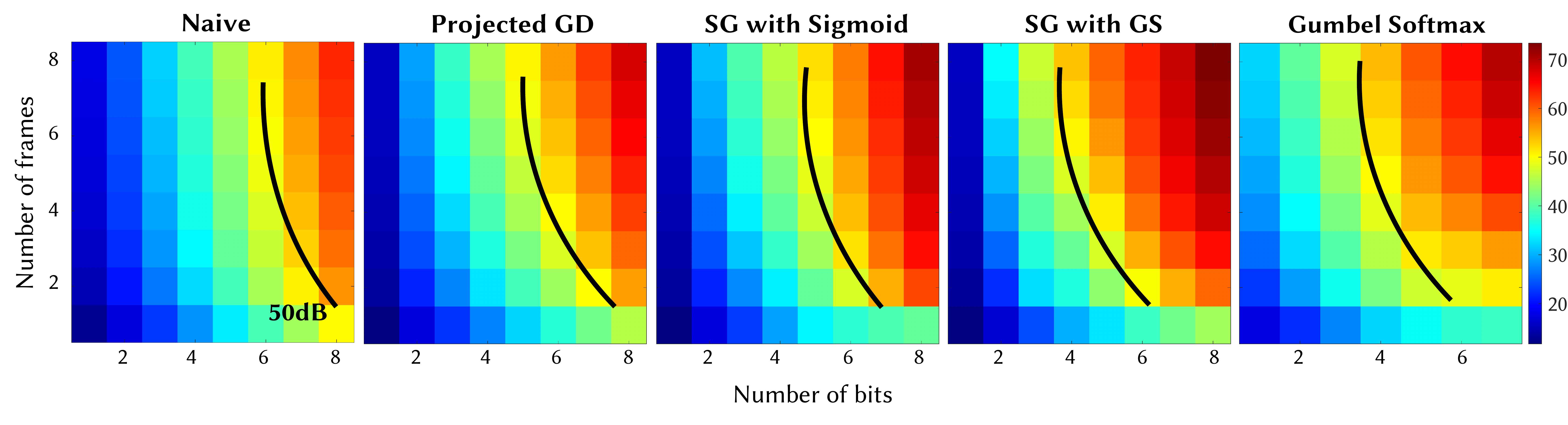}
		\caption{Trade-off between the number of frames and bits for quantized SLMs, using different optimization algorithms. We optimize SLM phase patterns with a different number of bits and frames for 14 target images. We simulate the setup using the ASM model for the green channel using 5 different methods, including the Naive approach that quantizes only at the end, a variant of the projected gradient descent approach that we elaborate in Sec.~S2.1, the Surrogate gradients approach with Sigmoid gradient, the Surrogate gradients with Gumbel-Softmax gradient, and the Gumbel-Softmax approach. We show averaged PSNR metrics as colormaps. In addition, we mark roughly where it reaches 50~dB PSNR as a black line, which is achieved by the 8~bit--1~frame type SLM. Note that the trend of black lines shifts.}
		\label{fig:simulation_tradeoff}
\end{figure*}

\paragraph{Exploring the trade-off between number of bits and frames}
We explore the trade-off space between the number of frames and the number of bits an SLM supports. We optimize phase patterns for 14 target images using the ASM model for the green channel using 5 different methods, including the Naive approach that quantizes only at the end, a variant of the projected gradient descent approach that we elaborate in Sec.~S2.1, the Surrogate gradients approach with Sigmoid gradient, the Surrogate gradients with Gumbel-Softmax gradient, and the Gumbel-Softmax approach. We show averaged PSNR metrics in colormaps in Fig.~\ref{fig:simulation_tradeoff}. 

Overall, the projected gradient descent approach improves upon the naive method, and notably, the surrogate gradient method with  the Gumbel-Softmax gradient outperforms other methods by a large margin. We also denote a line for each method that roughly matches 50~dB performance that the 8~bit--1~frame type SLM achieves. Accordingly, we observe the trend that advanced algorithms shift the line towards the bottom left, which means it can effectively save the number of bits and frames without sacrificing performance. Note that the last approach does not quantize during the optimization but only replaces the forward model with our forward model described in Eq.~6 in the manuscript. This is analogous to the Naive approach for the Surrogate gradients approach with Gumbel-Softmax. We note that this continuous relaxation is beneficial especially in more constrained cases.

\begin{figure*}[t!]
	\centering
		\includegraphics[width=1.95\columnwidth]{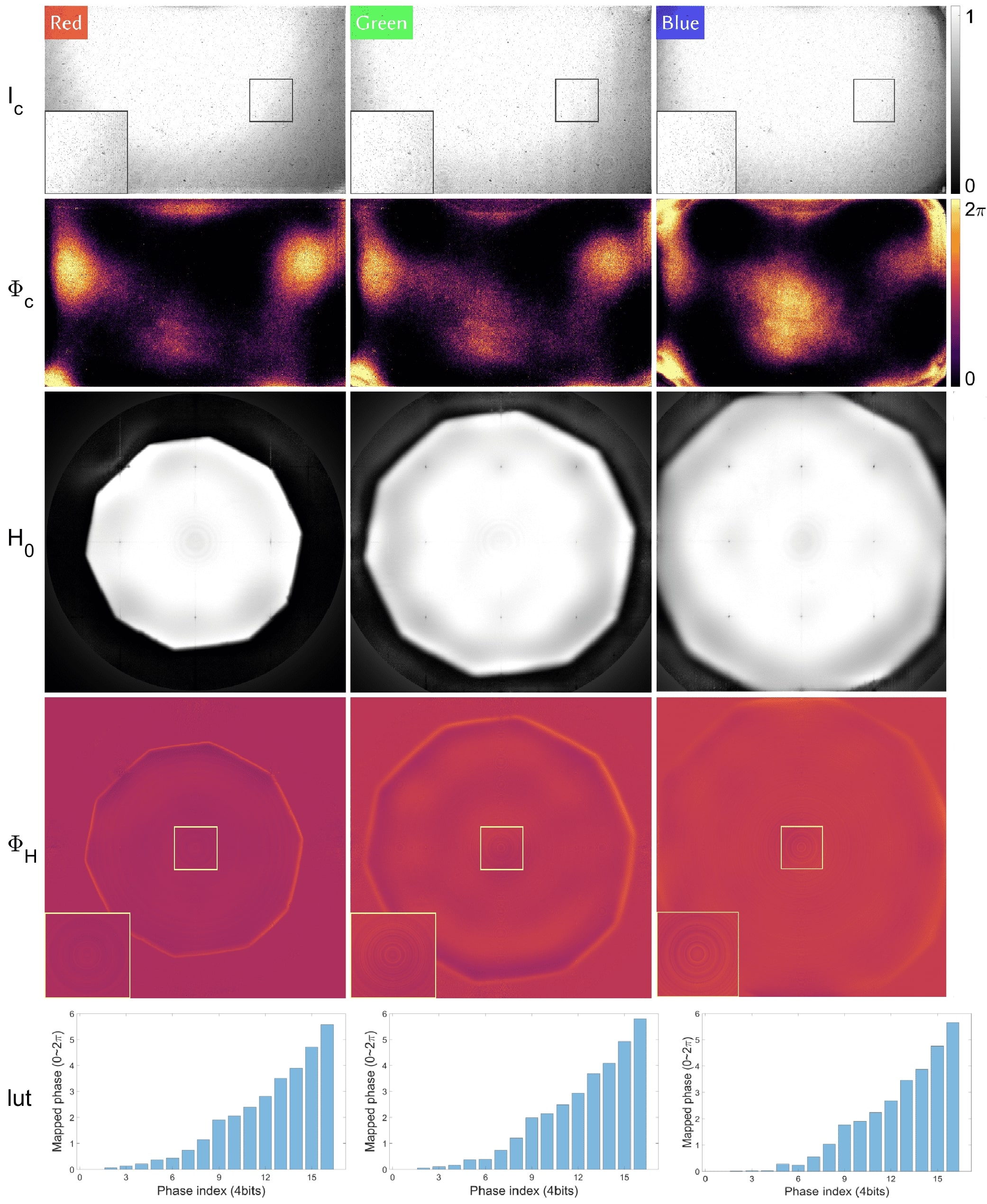}
		\caption{Parameters visualization of the calibrated model of our holographic display prototype (refer to Section~3 in main text). From left to right: red, green, and blue channels. From top to bottom: learned intensity on SLM plane, learned phase on SLM plane, learned amplitude of the optical filter on Fourier plane, learned phase of the optical filter on Fourier plane, and learned look up table for phase mapping.}
		\label{fig:model_vis}
\end{figure*}

\paragraph{Full model visualization}
Figure~\ref{fig:model_vis} visualizes our calibrated model, that includes learned intensity on SLM plane, learned phase on SLM plane, learned amplitude of the optical filter on Fourier plane, learned phase of the optical filter on Fourier plane, and learned lookup table for phase mapping. Note that this visualization includes many interesting aspects that present in the setup. First, the amplitude at SLM plane $a_{\textsc{SLM}}$ reveals the envelope of the incident beam as well as ripples and rings that occur in the physical display system. The phase at SLM plane $\phi_{\textsc{SLM}}$ shows the phase distortion. The terms at Fourier plane learn the shape of the physical filter we use in the setup and especially phase term $\phi_{\mathcal{F}}$ learns a radial phase ramp, and we note that potential propagation distance error can be learned through this parameter.

\begin{figure*}
\begin{adjustbox}{addcode={
\begin{minipage}[t]{\width}}{
\caption{Experimental holographic images reconstructed with different CGH algorithms when displaying one single frame and multiplexing 8 frames.}
\label{fig:model_results_compare_1}
\end{minipage}},rotate=90,center}
\includegraphics[width=1.2\textwidth]{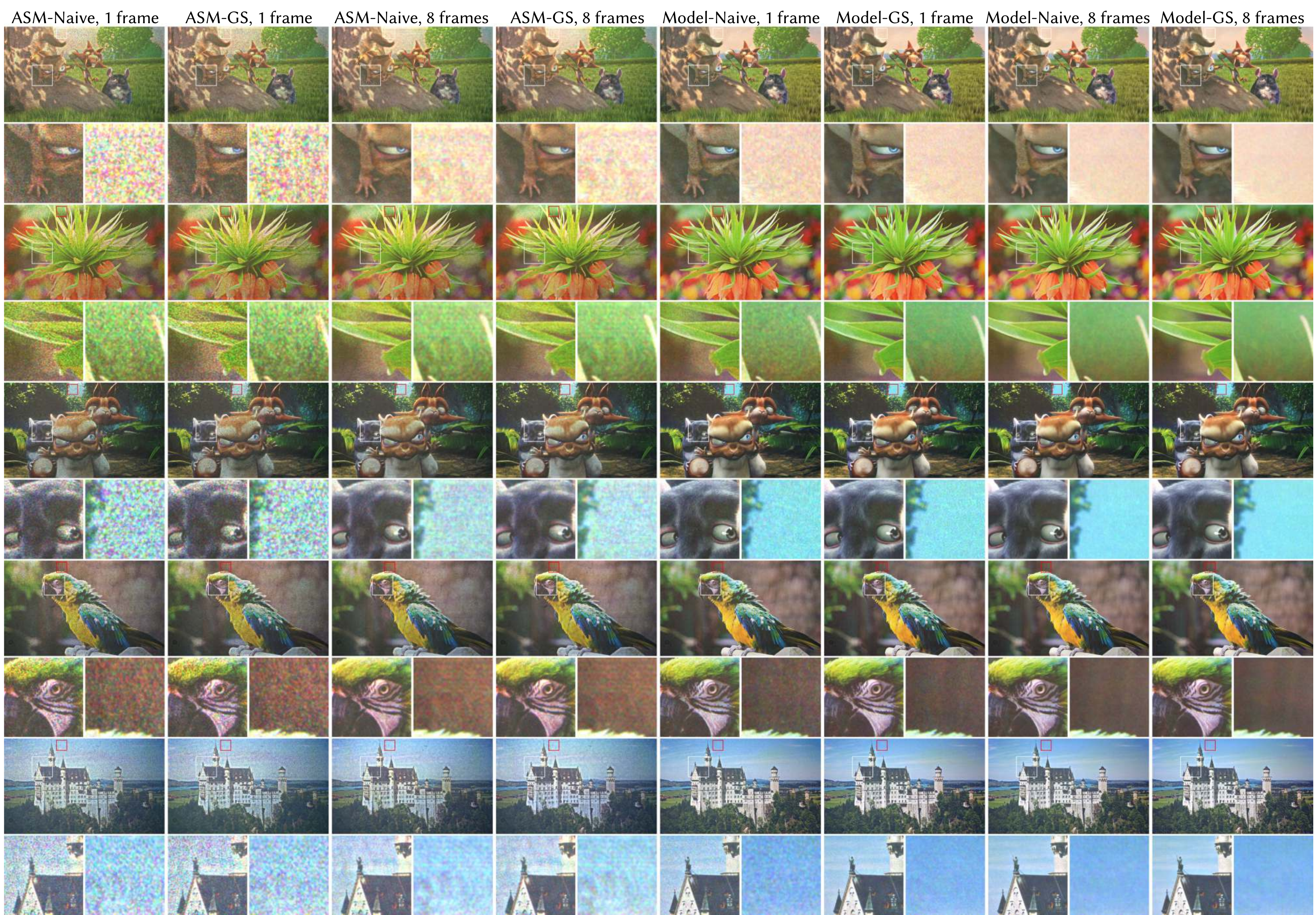}
\end{adjustbox}
\end{figure*}

\begin{figure*}
\begin{adjustbox}{addcode={
\begin{minipage}[t]{\width}}{
\caption{Experimental holographic images reconstructed with different CGH algorithms when displaying one single frame and multiplexing 8 frames.}
\label{fig:model_results_compare_2}
\end{minipage}},rotate=90,center}
\includegraphics[width=1.2\textwidth]{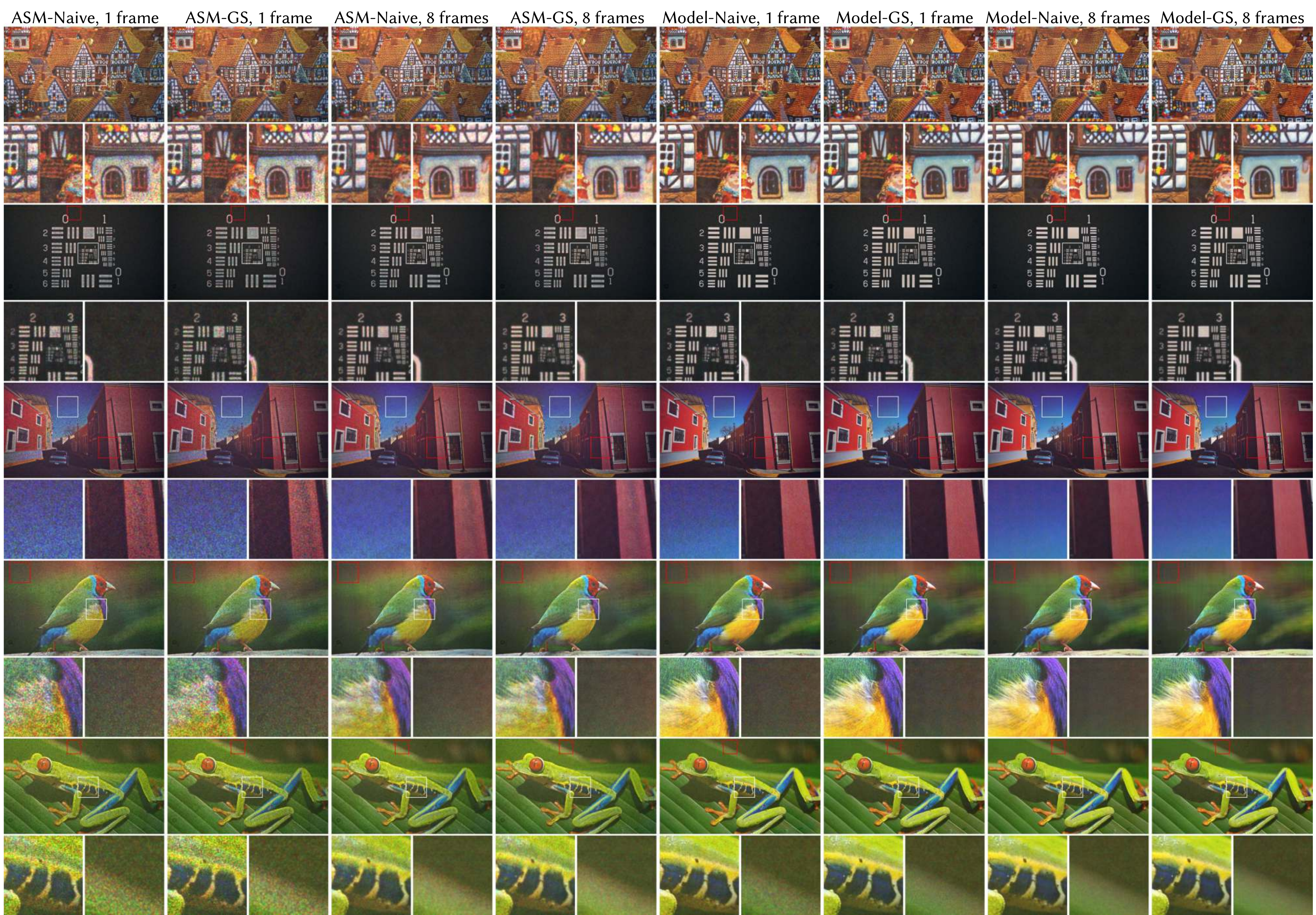}
\end{adjustbox}
\end{figure*}

\paragraph{Additional 2D results}
Figure~\ref{fig:model_results_compare_1} and Figure~\ref{fig:model_results_compare_2} present full resolution holographic images and their close-ups reconstructed with different CGH algorithms, including the ASM with naive quantization, the ASM with Gumbel-Softmax quantization, 
our Model with naive quantization, and our Model with Gumbel-Softmax quantization. For each algorithm, we assess with displaying one single frame on the SLM as well as multiplexing 8 frames (which is jointly optimized). Note that the benefits of Gumbel-Softmax are also less prominent in the ASM case when the image quality degradation is dominated by the model mismatch, but even then the quantitative evaluations indicate improved performance. Thus, the benefits of our quantization techniques are most significant when the model mismatch is mitigated using the learned model (See columns 5-8). Corresponding quantitative evaluation is presented in
Table~\ref{tab:resultscapturedPSNR8frs} and Table~\ref{tab:resultscapturedPSNR1frs}, indicating results with 8 multiplexed frames and 1 frame, respectively. PSNR and SSIM metrics are listed.

%

\begin{table*}[h]
	\caption{PSNR and SSIM metrics of captured 2D results with 8 multiplexed frames. Among all the methods, the proposed model, in tandem with the Gumbel-Softmax (GS) quantization, achieves the highest PSNR and SSIM. Images assessed are shown in Figure~\ref{fig:model_results_compare_1} (index 1 to 5) and Figure~\ref{fig:model_results_compare_2}  (index 6 to 10) .}
	\label{tab:resultscapturedPSNR8frs}	
	\begin{tabular}{l|ccccc}
		\toprule
		\rowcolor{gray!25}
		\multicolumn{5}{c}{Methods (algorithm-propagation operator)} \\ 
		\rowcolor{gray!25}		
		           &  SGD-ASM & SGD-ASM-GS & SGD-ours & SGD-ours-GS \\
		\midrule 
		\# 1     & 20.29 / 0.821 & 20.61 / 0.829 & 27.41 / 0.947 & 28.22 / 0.954 \\
		\# 2     & 17.24 / 0.796 & 17.52 / 0.807 & 22.39 / 0.909 & 23.00 / 0.916 \\
		\# 3     & 20.33 / 0.655 & 20.68 / 0.663 & 25.67 / 0.803 & 26.14 / 0.811 \\
		\# 4     & 18.43 / 0.632 & 18.63 / 0.643 & 22.35 / 0.815 & 22.70 / 0.811 \\
		\# 5     & 16.57 / 0.508 & 16.52 / 0.486 & 19.78 / 0.731 & 19.90 / 0.718 \\
		\# 6     & 18.26 / 0.789 & 18.48 / 0.799 & 23.33 / 0.911 & 23.82 / 0.915 \\
		\# 7     & 15.16 / 0.066 & 15.08 / 0.063 & 16.82 / 0.088 & 16.74/ 0.088 \\
		\# 8     & 18.09 / 0.654 & 18.15 / 0.643 & 21.23 / 0.764 & 21.24 / 0.758 \\
		\# 9     & 18.54 / 0.748 & 18.86 / 0.751 & 22.97 / 0.934 & 23.31 / 0.832 \\
		\# 10   & 18.62 / 0.820 & 18.83 / 0.827 & 23.09 / 0.889 & 23.50 / 0.898 \\
		\bottomrule
		\rowcolor{gray!25}
		Avg.     & 18.16 / 0.649 & 18.33 / 0.652 & 22.51/ 0.769 & 22.85 / 0.770 \\
		\bottomrule
	\end{tabular}
\end{table*}%

\begin{table*}[h]
	\caption{PSNR and SSIM metrics of captured 2D results with 1 single frame. Among all the methods, the proposed model, in tandem with the Gumbel-Softmax (GS) quantization, achieves the highest PSNR and SSIM. Images assessed are shown in Figure~\ref{fig:model_results_compare_1} (index 1 to 5) and Figure~\ref{fig:model_results_compare_2}  (index 6 to 10) .}
	\label{tab:resultscapturedPSNR1frs}	
	\begin{tabular}{l|ccccc}
		\toprule
		\rowcolor{gray!25}
		\multicolumn{5}{c}{Methods (algorithm-propagation operator)} \\ 
		\rowcolor{gray!25}		
		           &  SGD-ASM & SGD-ASM-GS & SGD-ours & SGD-ours-GS \\
		\midrule 
		\# 1     & 18.63 / 0.686 & 18.85 / 0.703 & 25.14 / 0.882 & 26.42 / 0.910 \\
		\# 2     & 16.31 / 0.722 & 16.30 / 0.721 & 20.89 / 0.864 & 22.30 / 0.888 \\
		\# 3     & 18.80 / 0.511 & 18.93 / 0.512 & 24.09 / 0.720 & 24.72 / 0.744 \\
		\# 4     & 17.31 / 0.480 & 17.38 / 0.481 & 21.26 / 0.709 & 21.75 / 0.715 \\
		\# 5     & 15.46 / 0.391 & 15.54 / 0.394 & 19.03 / 0.673 & 19.40 / 0.676 \\
		\# 6     & 17.48 / 0.719 & 17.53 / 0.724 & 22.33 / 0.875 & 22.96 / 0.889 \\
		\# 7     & 14.87 / 0.057 & 14.53 / 0.052 & 16.68 / 0.082 & 16.53 / 0.080 \\
		\# 8     & 16.88 / 0.555 & 17.01 / 0.563 & 20.32 / 0.709 & 20.57 / 0.713 \\
		\# 9     & 17.55 / 0.632 & 17.57 / 0.642 & 21.75 / 0.771 & 22.27 / 0.783 \\
		\# 10   & 17.73 / 0.758 & 17.71 / 0.762 & 21.90 / 0.858 & 22.57 / 0.871 \\
		\bottomrule
		\rowcolor{gray!25}
		Avg.     & 17.10 / 0.551 & 17.13 / 0.556 & 21.34 / 0.714 & 21.95 / 0.727 \\
		\bottomrule
	\end{tabular}
\end{table*}%

\begin{figure*}
\begin{adjustbox}{addcode={
\begin{minipage}[t]{\width}}{
\caption{Simulated focal stacks reconstructed with different CGH algorithms on ideal quantized SLMs. 
		From left to right: AADPM from Shi et al.~\cite{shi2021towards}, Model ADMM from Choi et al.~\cite{choi2021neural3d}, SGD-RGBD from Choi et al.~\cite{choi2021neural3d}, Randomness Control from Yoo et al.~\cite{Yoo:21}, our STFT RGBD implementation, and our focal stack implementation.  Focused regions are highlighted with red squares, that from top to bottom, indicate far, center, and near distances.}
\label{fig:simulation_3d_results_1}
\end{minipage}},rotate=90,center}
\includegraphics[width=1.22\textwidth]{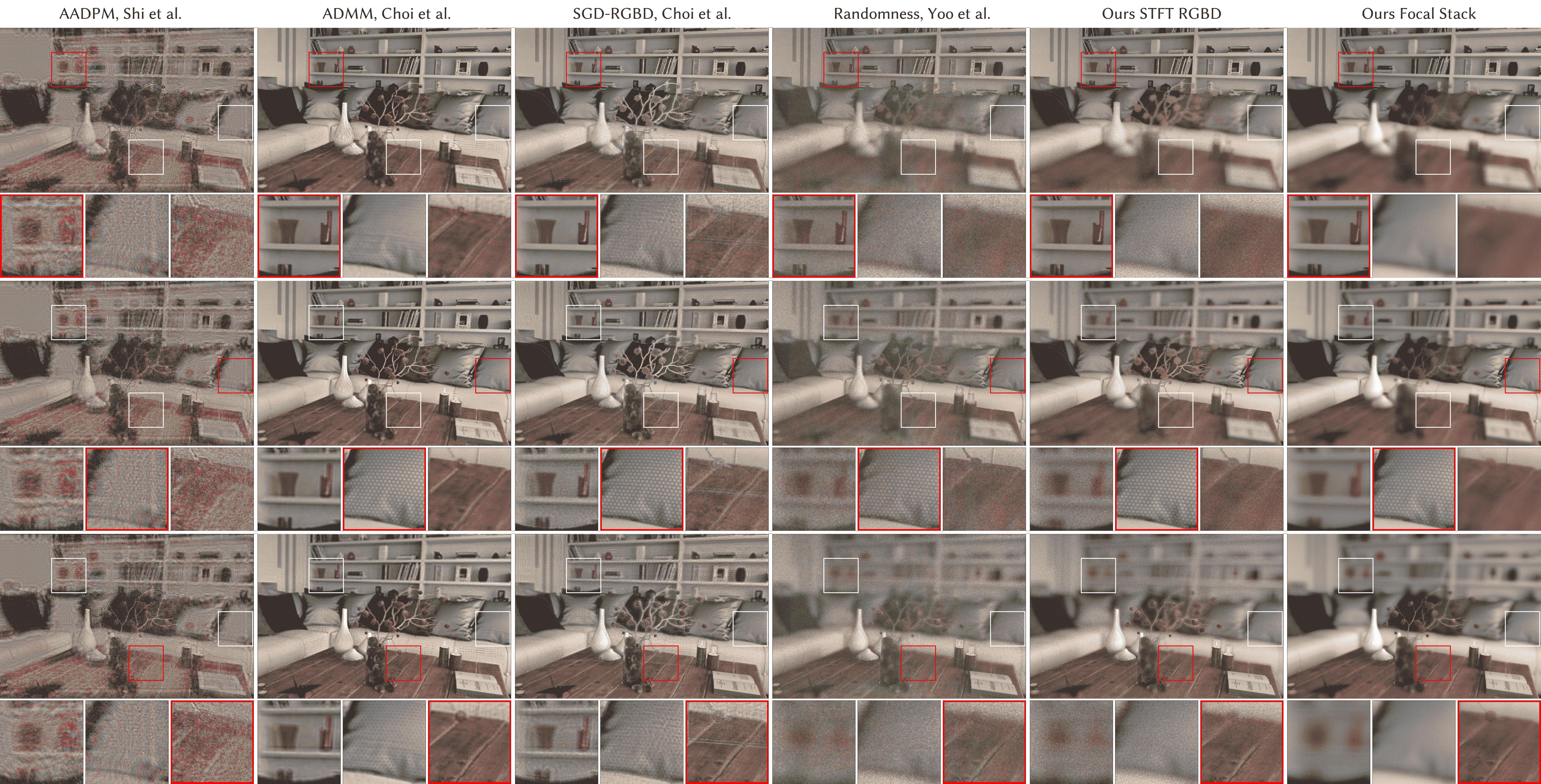}
\end{adjustbox}
\end{figure*}
\begin{figure*}
\begin{adjustbox}{addcode={
\begin{minipage}[t]{\width}}{
\caption{Simulated focal stacks reconstructed with different CGH algorithms on ideal quantized SLMs. 
		From left to right: AADPM from Shi et al.~\cite{shi2021towards}, Model ADMM from Choi et al.~\cite{choi2021neural3d}, SGD-RGBD from Choi et al.~\cite{choi2021neural3d}, Randomness Control from Yoo et al.~\cite{Yoo:21}, our STFT RGBD implementation, and our focal stack implementation.  Focused regions are highlighted with red squares, that from top to bottom, indicate far, center, and near distances.}
\label{fig:simulation_3d_results_2}
\end{minipage}},rotate=90,center}
\includegraphics[width=1.22\textwidth]{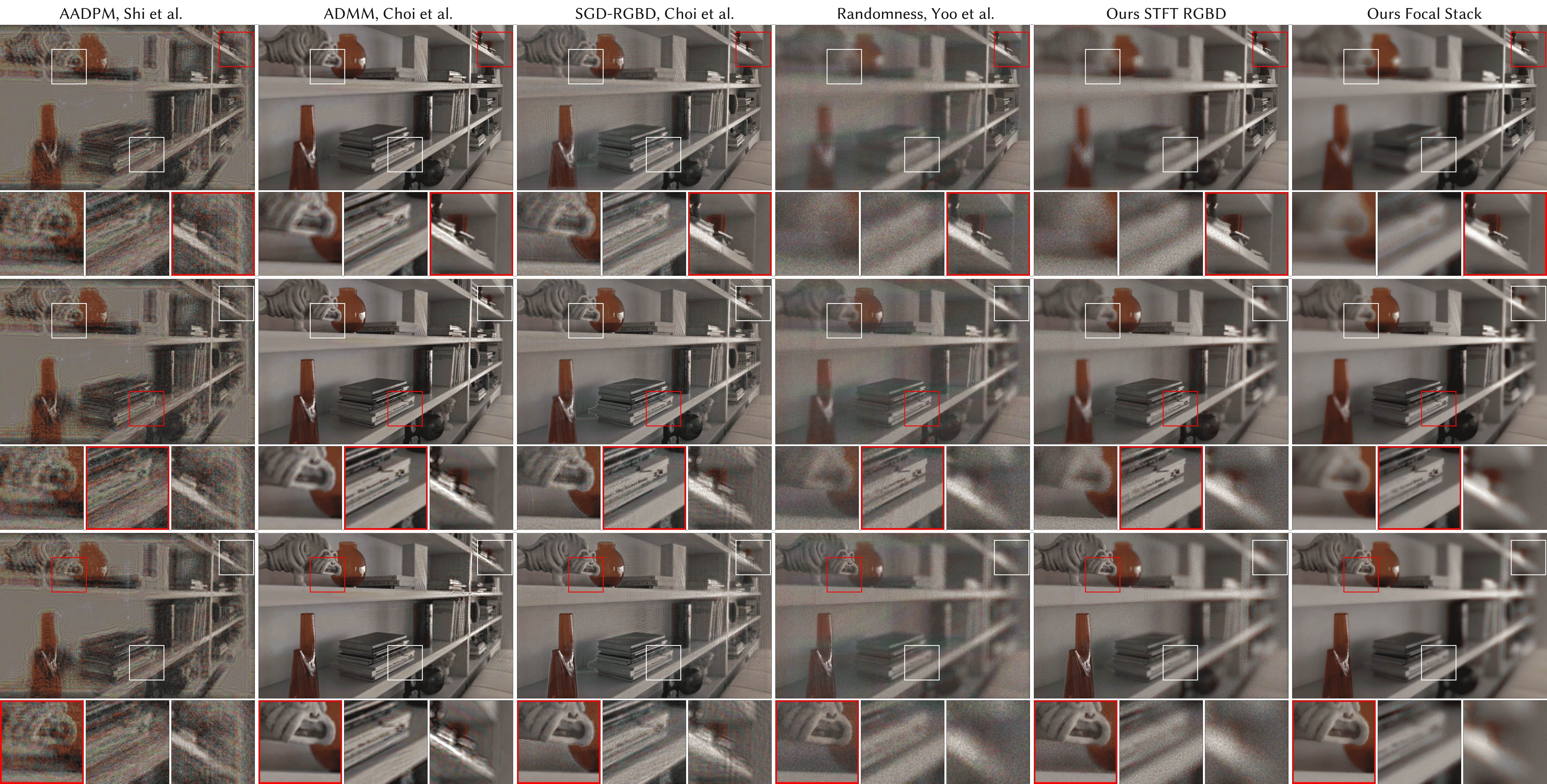}
\end{adjustbox}
\end{figure*}
\begin{figure*}
\begin{adjustbox}{addcode={
\begin{minipage}[t]{\width}}{
\caption{Experimentally captured focal stacks reconstructed with different CGH algorithms on ideal quantized SLMs. 
		From left to right: AADPM from Shi et al.~\cite{shi2021towards}, Model ADMM from Choi et al.~\cite{choi2021neural3d}, SGD-RGBD from Choi et al.~\cite{choi2021neural3d}, Randomness Control from Yoo et al.~\cite{Yoo:21}, our STFT RGBD implementation, and our focal stack implementation.  Focused regions are highlighted with red squares, that from top to bottom, indicate far, center, and near distances.}
\label{fig:simulation_3d_results_experiments_1}
\end{minipage}},rotate=90,center}
\includegraphics[width=1.22\textwidth]{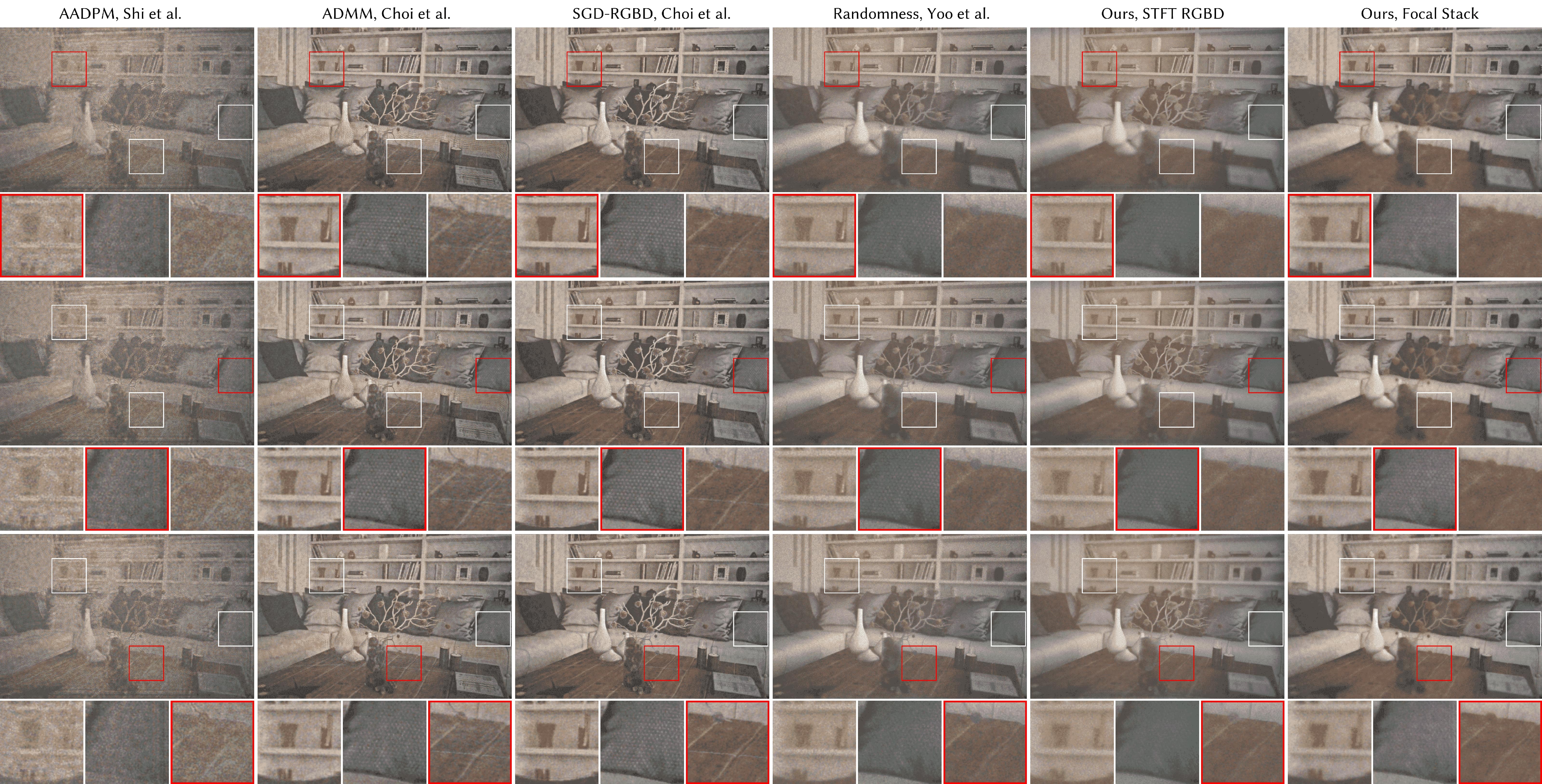}
\end{adjustbox}
\end{figure*}
\begin{figure*}
\begin{adjustbox}{addcode={
\begin{minipage}[t]{\width}}{
\caption{Experimentally captured  focal stacks reconstructed with different CGH algorithms on the Gumbel-Softmax (GS) quantized SLMs. 
		From left to right: AADPM from Shi et al.~\cite{shi2021towards}, Model ADMM from Choi et al.~\cite{choi2021neural3d}, SGD-RGBD from Choi et al.~\cite{choi2021neural3d}, Randomness Control from Yoo et al.~\cite{Yoo:21}, our STFT RGBD implementation, and our focal stack implementation. Note that ADMM with GS doesn't work in this case. Focused regions are highlighted with red squares, that from top to bottom, indicate far, center, and near distances.}
\label{fig:simulation_3d_results_experiments_2}
\end{minipage}},rotate=90,center}
\includegraphics[width=1.05\textwidth]{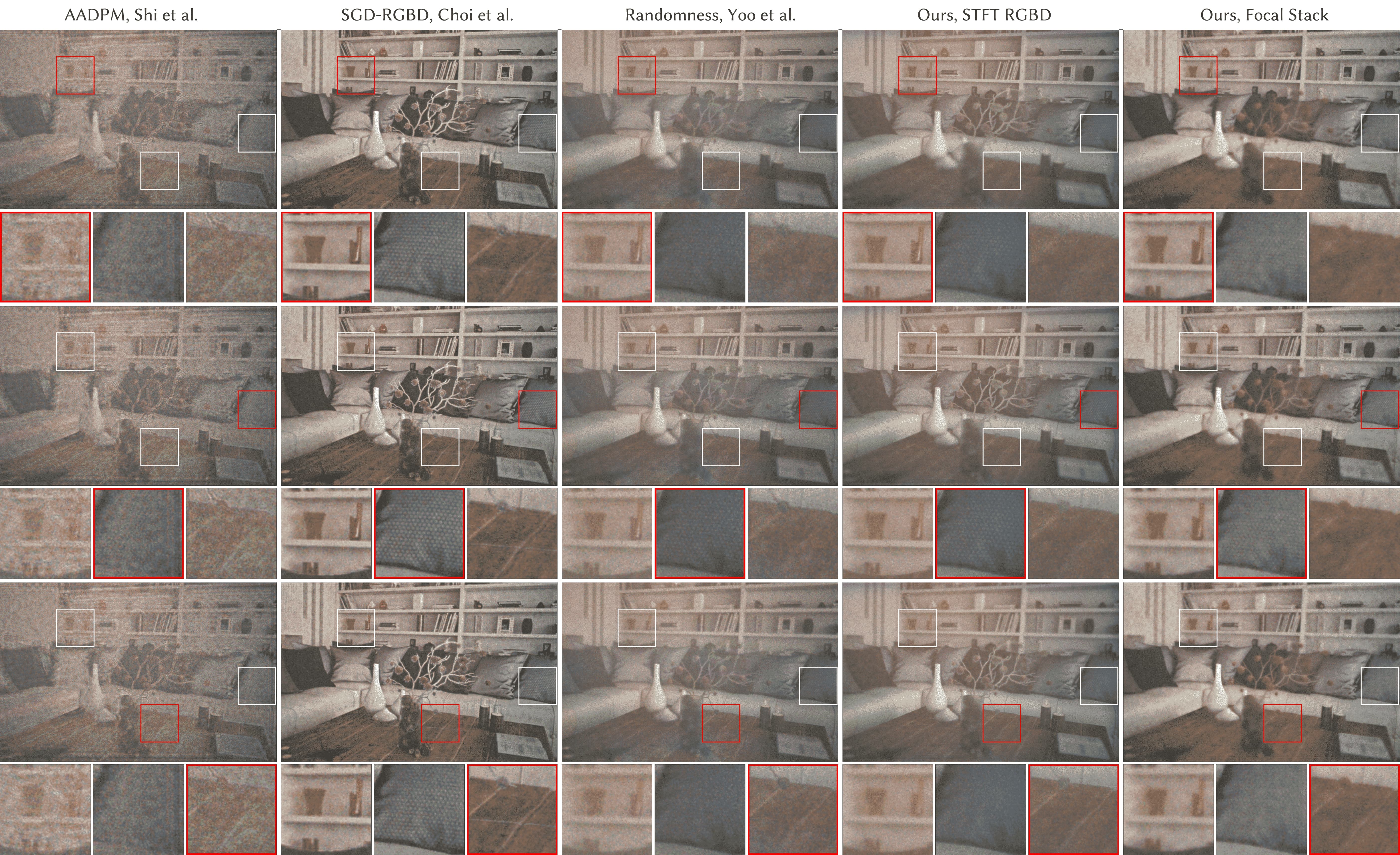}
\end{adjustbox}
\end{figure*}

\paragraph{Additional 3D results}
Figure~\ref{fig:simulation_3d_results_1} and Figure~\ref{fig:simulation_3d_results_2} further present comprehensive simulation results of focal stacks and their close-ups reconstructed with different CGH algorithms, including the AADPM from Shi et al.~\cite{shi2021towards}, the Model ADMM from Choi et al.~\cite{choi2021neural3d}, the SGD-RGBD from Choi et al.~\cite{choi2021neural3d}, the randomness control from Yoo et al.~\cite{Yoo:21}, our STFT RGBD implementation, and our focal stack implementation. All of these holograms used are with quantization operations. For each algorithm, reconstructed images with the camera focus at three different distances (far, center, near) are shown. We observe that ours outperform the others in preserving sharp content for in-focus regions while providing more natural blur for out-of-focus regions, with the focal stack implementation on the right being the best. Accordingly, we experimentally captured results of the scene in Figure~\ref{fig:simulation_3d_results_1} optimized with native quantization and Gumbel-Softmax (GS) quantization, as shown in Figure~\ref{fig:simulation_3d_results_experiments_1} and Figure~\ref{fig:simulation_3d_results_experiments_2}.

\begin{figure*}[t!]
	\centering
		\includegraphics[width=2.1\columnwidth]{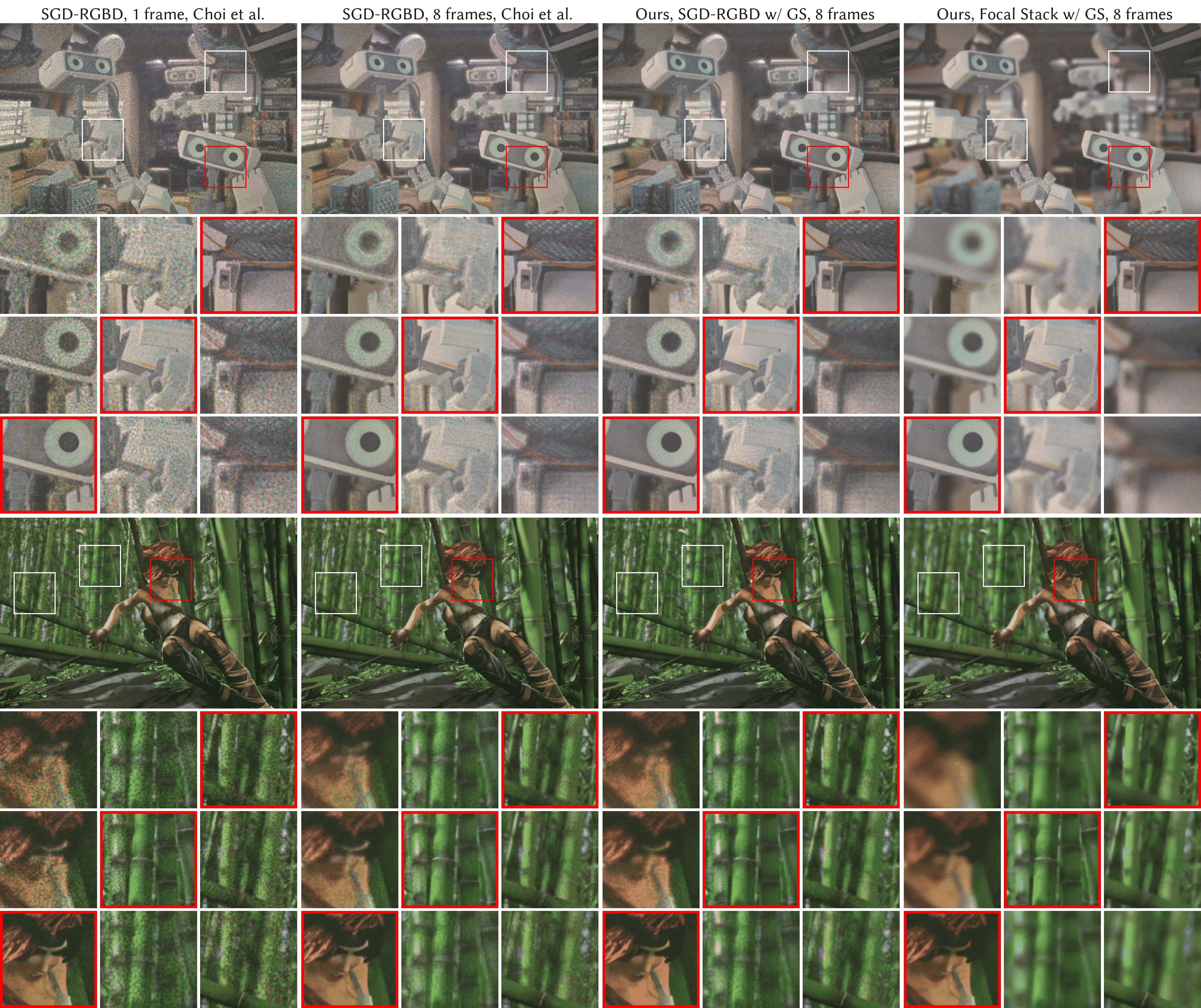}
		\vspace{-6pt}
		\caption{Comparison of 3D CGH algorithms using experimentally captured data. In this experiment, we compare algorithms of SGD-NH3D using RGBD input~\cite{choi2021neural3d} with 1 frame and 8 multiplexed frames, respectively, SGD-ours using RGBD input with Gumbel-Softmax (GS), and SGD-ours using Focal Stack with GS. For the top scene from left to right, the corresponding PSNR metrics are 23.5~dB, 25.6~dB, 28.7~dB, 27.7~dB with respect to the RGBD all-in-focus targets, and 19.7~dB, 21.6~dB, 23.6~dB, 26.1~dB with respect to the focal stack. Same metrics for the bottom scene are 28.3~dB, 30.0~dB, 31.0~dB, 30.3~dB and 24.3~dB, 26.5~dB, 27.0~dB, 30.1~dB.
For close-ups, red squares indicate where the camera is focused at three distances (from top to bottom: far, intermediate, and near). }
		\label{fig:experiment_results_3D_1}
\end{figure*}

\begin{figure*}[t!]
	\centering
		\includegraphics[width=2.1\columnwidth]{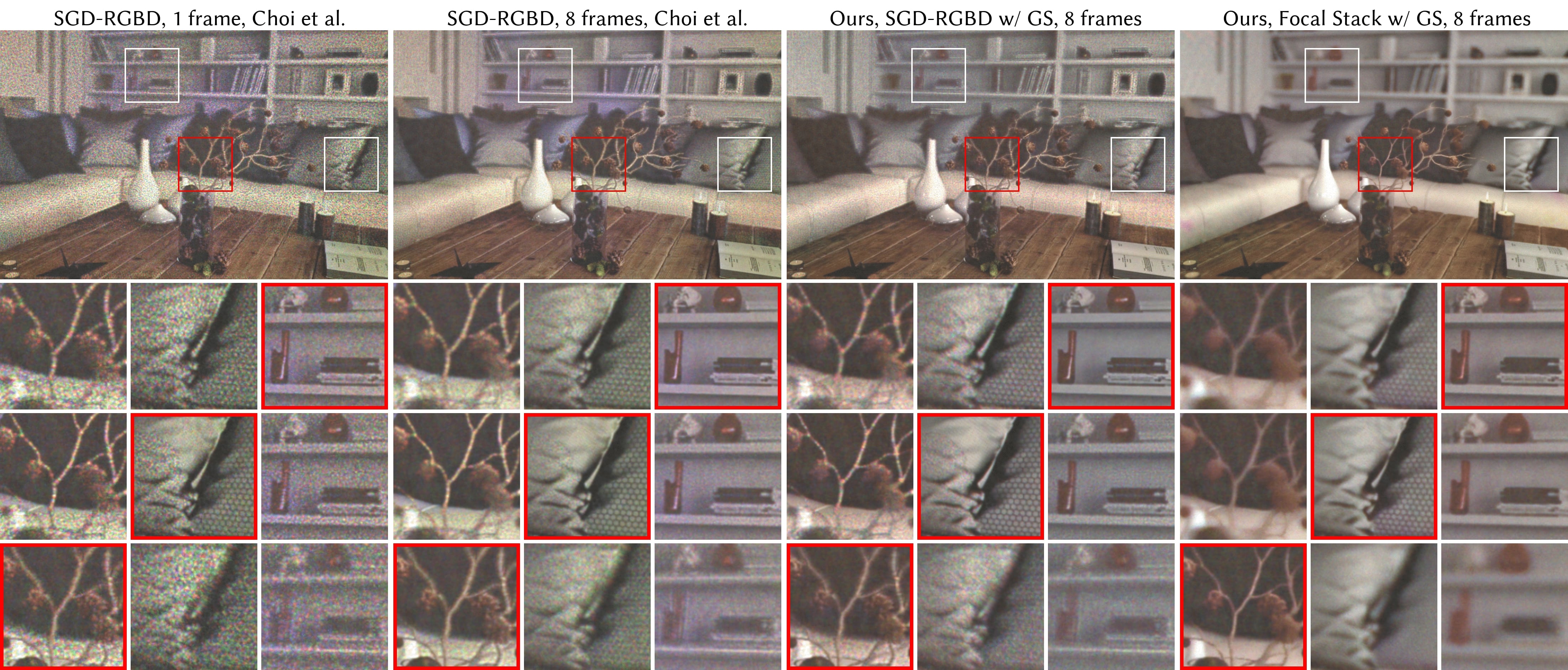}
		\vspace{-6pt}
		\caption{Comparison of 3D CGH algorithms using experimentally captured data. In this experiment, we compare algorithms of SGD-NH3D using RGBD input~\cite{choi2021neural3d} with 1 frame and 8 multiplexed frames, respectively, our SGD-RGBD with Gumbel-Softmax (GS), and our SGD-Focal Stack with GS. From left to right, the corresponding PSNR metrics are 21.2~dB, 22.5~dB, 24.7~dB, 24.2~dB with respect to the RGBD all-in-focus targets, and 18.5~dB, 20.3~dB, 22.3~dB, 23.7~dB with respect to the focal stack.
For close-ups, red squares indicate where the camera is focused at three distances (from top to bottom: far, intermediate, and near). }
		\label{fig:experiment_results_3D_2}
\end{figure*}

\begin{figure*}[t!]
	\centering
		\includegraphics[width=1.23\columnwidth]{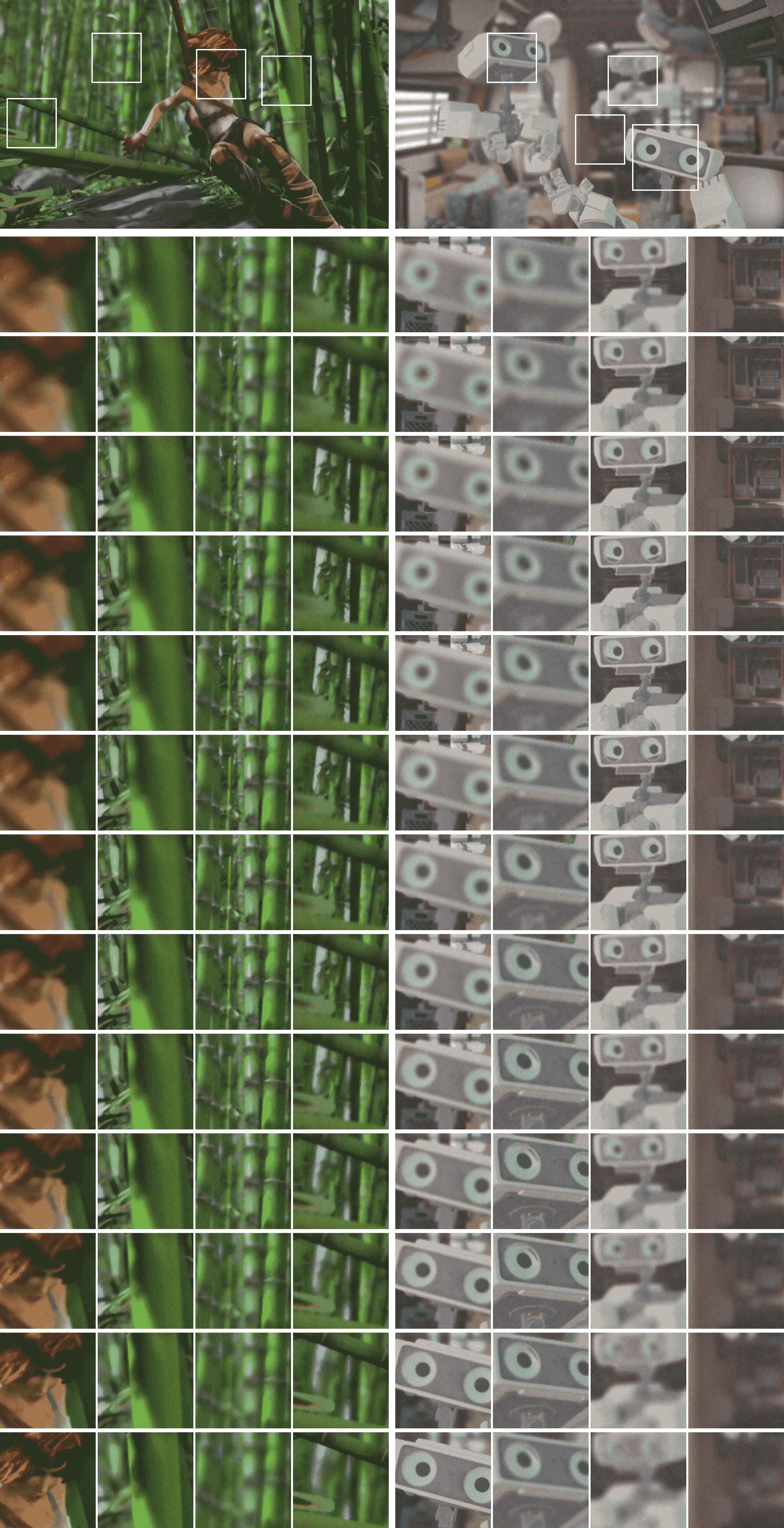}
		\vspace{-6pt}
		\caption{Interpolation behaviour of 3D focal stack supervised holograms. We experimentally capture 13 planes and show them at each row of closeups. Note that only odd rows are supervised while unsupervised planes (even rows) interpolate it smoothly.}
		\label{fig:experiment_results_3D_interp}
\end{figure*}

Figure~\ref{fig:experiment_results_3D_1} and Figure~\ref{fig:experiment_results_3D_2} show additional experimental results of 3D holographic display assessing different CGH algorithms (complimentary to Figure~6 in the main paper).
In this experiment, we compare algorithms of the SGD-NH3D using RGBD input~\cite{choi2021neural3d} with 1 frame and 8 multiplexed frames, respectively, SGD-ours using RGBD input without and with Gumbel-Softmax (GS), and SGD-ours using Focal Stack with GS. Quantitative assessments are provided as PSNR metrics in the caption, as well as summarized in Table~\ref{tab:resultscapturedPSNR_3d}. We also show the behaviour of the interpolation between supervised planes in Fig.~\ref{fig:experiment_results_3D_interp}.

\begin{figure*}
\begin{adjustbox}{addcode={
\begin{minipage}[t]{\width}}{
\caption{Comparison of 4D CGH algorithms using experimentally captured data. In this experiment, we compare the OLAS algorithm~\cite{Padmanaban:2019}, and the SGD algorithms using our ASM-Naive with 1 single frame, our ASM-GS with 1 single frame, our ASM-Naive with 8 multiplexed frames, our ASM-GS with 8 multiplexed frames, and our Model-GS with 8 multiplexed frames. For close-ups, red squares indicate where the camera is focused at three distances (from top to bottom: far, intermediate, and near). }
\label{fig:results_4D}
\end{minipage}},rotate=90,center}
\includegraphics[width=1.15\textwidth]{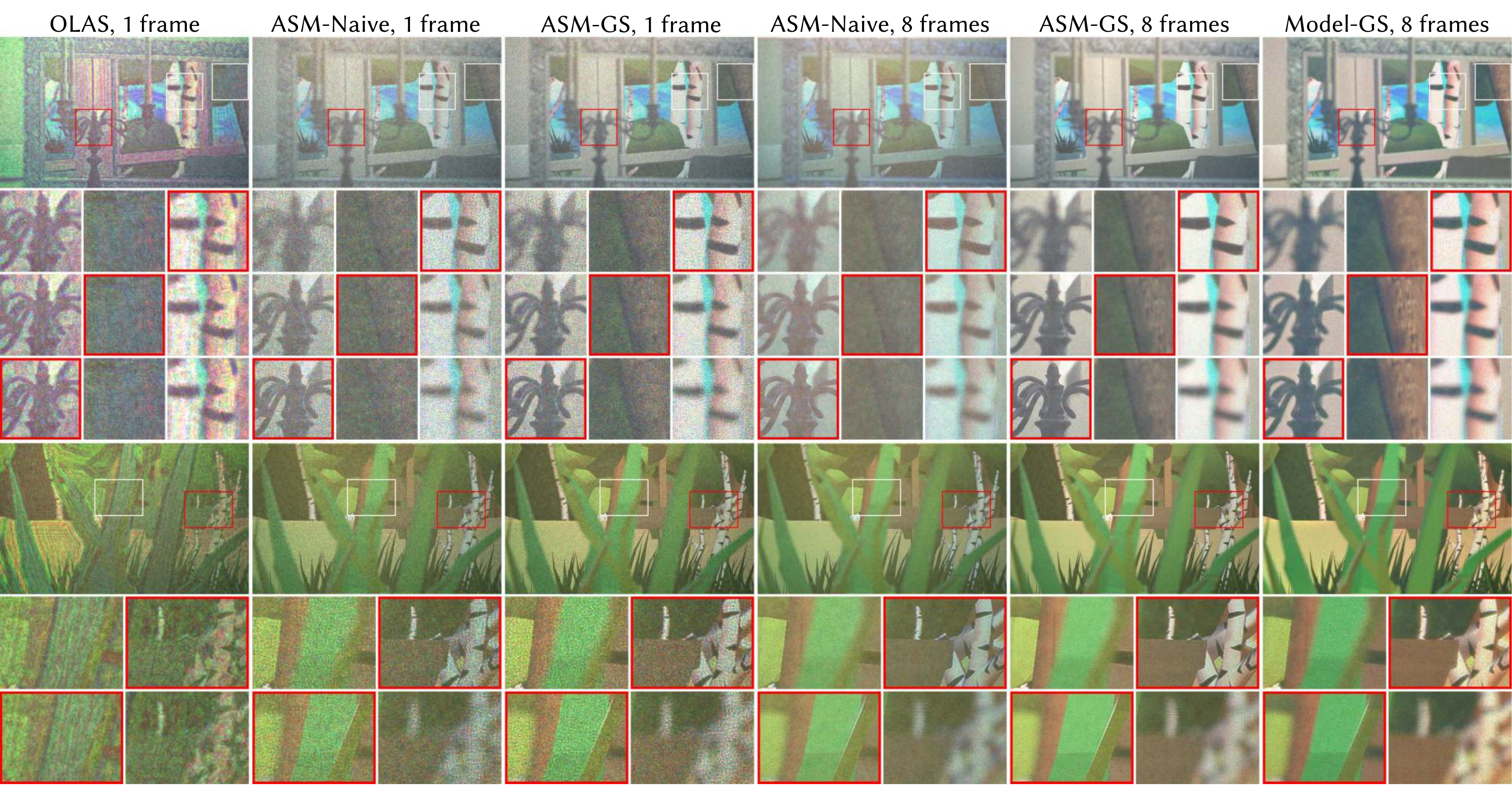}
\end{adjustbox}
\end{figure*}

\paragraph{Additional 4D results}
Figure~\ref{fig:results_4D} presents experimental results of light field reconstructed with different CGH algorithms, including the ASM-Naive with 1 single frame, our ASM-GS with 1 single frame, the ASM-Naive with 8 multiplexed frames, and our ASM-GS with 8 multiplexed frames. For each example scene, we show close-ups of content at three distances (far, intermediate, near). Our framework leads to overall higher image fidelity for both the in-focus and out-of-focus regions.

\begin{figure*}[t!]
	\centering
		\includegraphics[width=2.1\columnwidth]{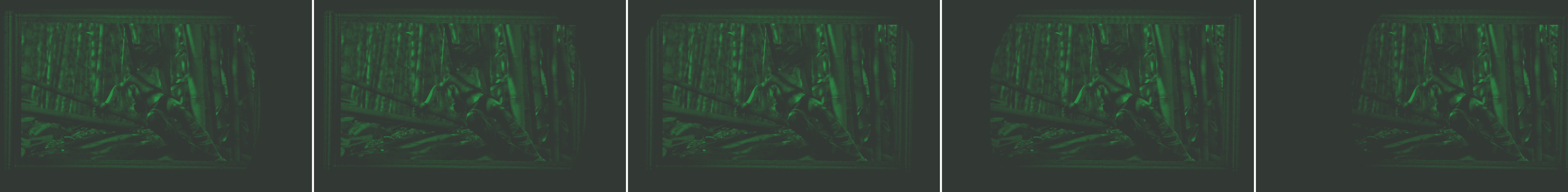}
		\caption{Frames extracted from the camera with different spatial shifts.}
		\label{fig:viewpoints}
\end{figure*}

\paragraph{Robustness to possible viewpoint Shifts}
Figure~\ref{fig:viewpoints} presents a set of captured results of a holographic scene that validates the robustness of our image synthesis to possible viewpoint shifts. The camera is manually translated in horizontal from left to right, for a few millimeters. We observe no noticeable degradation in image quality over the viewpoint shifts.
%
%

\newpage

\newpage

\newpage

\newpage

\begin{table*}[h]
	\caption{PSNR metrics of captured 3D results using different CGH algorithms, including the SGD-NH3D using RGBD input~\cite{choi2021neural3d} with 1 frame and 8 multiplexed frames, respectively, SGD-our model using RGBD input, SGD-our model using RGBD input with Gumbel-Softmax (GS), and SGD-our model using Focal Stack (FS) supervision with GS. Images assessed are shown in Figure~\ref{fig:experiment_results_3D_1}, Figure~\ref{fig:experiment_results_3D_2}, and Figure~6 in the main paper. For each cell, the first PSNR is evaluated with respect to the RGBD all-in-focus targets, while the second one with respect to the focal stack. Note that the first four columns are supervised on RGBD input, where ours achieves the best all-in-focus PSNR, and the fifth column is supervised on a focal stack, and achieves the best performance on the PSNR metric on the target focal stack.}
	\label{tab:resultscapturedPSNR_3d}	
	\begin{tabular}{l|cccccc}
		\toprule
		\rowcolor{gray!25}
		\multicolumn{6}{c}{Methods (SGD-propagation operator)} \\ 
		\rowcolor{gray!25}		
		           &  NH3D, 1 frame & NH3D, 8 frames & ours, 8 frames & ours,  w/ GS, 8 frames & ours, w/ GS (FS), 8 frames \\
		\midrule 
		\bf Robot                   & 23.5 / 19.7 & 25.6 / 21.6 & 27.9 / 23.2 & {\bf 28.7} / 23.6 & 27.7 / {\bf 26.1} \\
		\bf Sintel Bamboo     & 28.3 / 24.3 & 30.0 / 26.5 & 30.4 / 26.9 & {\bf 31.0} / 27.0 & 30.3 / {\bf 30.1} \\
		\bf Hyperism Room   & 21.2 / 18.5 & 22.5 / 20.3 & 24.0 / 21.7 &{\bf 24.7} / 22.3 & 24.2 / {\bf 23.7} \\
		\bf Big Buck Bunny    & 24.3 / 21.3 & 25.8 / 23.2 & 26.1 / 24.0 & {\bf 26.7} / 24.5 & 25.9 / {\bf 26.9} \\
		\bottomrule
	\end{tabular}
\end{table*}%

%% file: ms.bbl

\begin{thebibliography}{48}


\ifx \showCODEN    \undefined \def \showCODEN     #1{\unskip}     \fi
\ifx \showDOI      \undefined \def \showDOI       #1{#1}\fi
\ifx \showISBNx    \undefined \def \showISBNx     #1{\unskip}     \fi
\ifx \showISBNxiii \undefined \def \showISBNxiii  #1{\unskip}     \fi
\ifx \showISSN     \undefined \def \showISSN      #1{\unskip}     \fi
\ifx \showLCCN     \undefined \def \showLCCN      #1{\unskip}     \fi
\ifx \shownote     \undefined \def \shownote      #1{#1}          \fi
\ifx \showarticletitle \undefined \def \showarticletitle #1{#1}   \fi
\ifx \showURL      \undefined \def \showURL       {\relax}        \fi
\providecommand\bibfield[2]{#2}
\providecommand\bibinfo[2]{#2}
\providecommand\natexlab[1]{#1}
\providecommand\showeprint[2][]{arXiv:#2}

\bibitem[\protect\citeauthoryear{Bartlett, McDonald, and Hall}{Bartlett
  et~al\mbox{.}}{2019}]%
        {Bartlett:2019}
\bibfield{author}{\bibinfo{person}{Terry~A. Bartlett},
  \bibinfo{person}{William~C. McDonald}, {and} \bibinfo{person}{James~N.
  Hall}.} \bibinfo{year}{2019}\natexlab{}.
\newblock \showarticletitle{Adapting Texas Instruments DLP technology to
  demonstrate a phase spatial light modulator}. In
  \bibinfo{booktitle}{\emph{SPIE OPTO, Proceedings Volume 10932, Emerging
  Digital Micromirror Device Based Systems and Applications XI}}.
  \bibinfo{pages}{109320S}.
\newblock


\bibitem[\protect\citeauthoryear{Bengio, L{\'e}onard, and Courville}{Bengio
  et~al\mbox{.}}{2013}]%
        {bengio2013estimating}
\bibfield{author}{\bibinfo{person}{Yoshua Bengio}, \bibinfo{person}{Nicholas
  L{\'e}onard}, {and} \bibinfo{person}{Aaron Courville}.}
  \bibinfo{year}{2013}\natexlab{}.
\newblock \showarticletitle{Estimating or propagating gradients through
  stochastic neurons for conditional computation}.
\newblock \bibinfo{journal}{\emph{arXiv preprint arXiv:1308.3432}}
  (\bibinfo{year}{2013}).
\newblock


\bibitem[\protect\citeauthoryear{Benton}{Benton}{1983}]%
        {Benton:1983}
\bibfield{author}{\bibinfo{person}{Stephen~A. Benton}.}
  \bibinfo{year}{1983}\natexlab{}.
\newblock \showarticletitle{Survey Of Holographic Stereograms}. In
  \bibinfo{booktitle}{\emph{Proc. SPIE}}, Vol.~\bibinfo{volume}{0367}.
\newblock


\bibitem[\protect\citeauthoryear{Boyd, Boyd, and Vandenberghe}{Boyd
  et~al\mbox{.}}{2004}]%
        {boyd2004convex}
\bibfield{author}{\bibinfo{person}{Stephen Boyd}, \bibinfo{person}{Stephen~P
  Boyd}, {and} \bibinfo{person}{Lieven Vandenberghe}.}
  \bibinfo{year}{2004}\natexlab{}.
\newblock \bibinfo{booktitle}{\emph{Convex optimization}}.
\newblock \bibinfo{publisher}{Cambridge university press}.
\newblock


\bibitem[\protect\citeauthoryear{Chakravarthula, Peng, Kollin, Fuchs, and
  Heide}{Chakravarthula et~al\mbox{.}}{2019}]%
        {Chakravarthula:2019}
\bibfield{author}{\bibinfo{person}{Praneeth Chakravarthula},
  \bibinfo{person}{Yifan Peng}, \bibinfo{person}{Joel Kollin},
  \bibinfo{person}{Henry Fuchs}, {and} \bibinfo{person}{Felix Heide}.}
  \bibinfo{year}{2019}\natexlab{}.
\newblock \showarticletitle{Wirtinger Holography for Near-eye Displays}.
\newblock \bibinfo{journal}{\emph{ACM Trans. Graph.}} \bibinfo{volume}{38},
  \bibinfo{number}{6} (\bibinfo{year}{2019}).
\newblock


\bibitem[\protect\citeauthoryear{Chakravarthula, Tseng, Srivastava, Fuchs, and
  Heide}{Chakravarthula et~al\mbox{.}}{2020}]%
        {chakravarthula2020learned}
\bibfield{author}{\bibinfo{person}{Praneeth Chakravarthula},
  \bibinfo{person}{Ethan Tseng}, \bibinfo{person}{Tarun Srivastava},
  \bibinfo{person}{Henry Fuchs}, {and} \bibinfo{person}{Felix Heide}.}
  \bibinfo{year}{2020}\natexlab{}.
\newblock \showarticletitle{Learned hardware-in-the-loop phase retrieval for
  holographic near-eye displays}.
\newblock \bibinfo{journal}{\emph{ACM Trans. on Graph. (TOG)}}
  \bibinfo{volume}{39}, \bibinfo{number}{6} (\bibinfo{year}{2020}),
  \bibinfo{pages}{1--18}.
\newblock


\bibitem[\protect\citeauthoryear{Chang, Bang, Wetzstein, Lee, and Gao}{Chang
  et~al\mbox{.}}{2020}]%
        {Chang:20}
\bibfield{author}{\bibinfo{person}{Chenliang Chang}, \bibinfo{person}{Kiseung
  Bang}, \bibinfo{person}{Gordon Wetzstein}, \bibinfo{person}{Byoungho Lee},
  {and} \bibinfo{person}{Liang Gao}.} \bibinfo{year}{2020}\natexlab{}.
\newblock \showarticletitle{Toward the next-generation VR/AR optics: a review
  of holographic near-eye displays from a human-centric perspective}.
\newblock \bibinfo{journal}{\emph{Optica}} \bibinfo{volume}{7},
  \bibinfo{number}{11} (\bibinfo{year}{2020}), \bibinfo{pages}{1563--1578}.
\newblock


\bibitem[\protect\citeauthoryear{Chen, Lee, Li, Chae, Wang, Wang, and Lee}{Chen
  et~al\mbox{.}}{2021}]%
        {Chen:21}
\bibfield{author}{\bibinfo{person}{Chun Chen}, \bibinfo{person}{Byounghyo Lee},
  \bibinfo{person}{Nan-Nan Li}, \bibinfo{person}{Minseok Chae},
  \bibinfo{person}{Di Wang}, \bibinfo{person}{Qiong-Hua Wang}, {and}
  \bibinfo{person}{Byoungho Lee}.} \bibinfo{year}{2021}\natexlab{}.
\newblock \showarticletitle{Multi-depth hologram generation using stochastic
  gradient descent algorithm with complex loss function}.
\newblock \bibinfo{journal}{\emph{Opt. Express}} \bibinfo{volume}{29},
  \bibinfo{number}{10} (\bibinfo{year}{2021}), \bibinfo{pages}{15089--15103}.
\newblock


\bibitem[\protect\citeauthoryear{Chen and Chu}{Chen and Chu}{2015}]%
        {Chen:15}
\bibfield{author}{\bibinfo{person}{Jhen-Si Chen} {and} \bibinfo{person}{Daping
  Chu}.} \bibinfo{year}{2015}\natexlab{}.
\newblock \showarticletitle{Improved layer-based method for rapid hologram
  generation and real-time interactive holographic display applications}.
\newblock \bibinfo{journal}{\emph{Opt. Express}} \bibinfo{volume}{23},
  \bibinfo{number}{14} (\bibinfo{year}{2015}), \bibinfo{pages}{18143--18155}.
\newblock


\bibitem[\protect\citeauthoryear{Chen and Wilkinson}{Chen and
  Wilkinson}{2009}]%
        {Chen:2009}
\bibfield{author}{\bibinfo{person}{Rick H-Y Chen} {and}
  \bibinfo{person}{Timothy~D Wilkinson}.} \bibinfo{year}{2009}\natexlab{}.
\newblock \showarticletitle{Computer generated hologram with geometric
  occlusion using GPU-accelerated depth buffer rasterization for
  three-dimensional display}.
\newblock \bibinfo{journal}{\emph{Applied optics}} \bibinfo{volume}{48},
  \bibinfo{number}{21} (\bibinfo{year}{2009}), \bibinfo{pages}{4246--4255}.
\newblock


\bibitem[\protect\citeauthoryear{Choi, Gopakumar, Peng, Kim, and
  Wetzstein}{Choi et~al\mbox{.}}{2021a}]%
        {choi2021neural3d}
\bibfield{author}{\bibinfo{person}{Suyeon Choi}, \bibinfo{person}{Manu
  Gopakumar}, \bibinfo{person}{Yifan Peng}, \bibinfo{person}{Jonghyun Kim},
  {and} \bibinfo{person}{Gordon Wetzstein}.} \bibinfo{year}{2021}\natexlab{a}.
\newblock \showarticletitle{Neural 3D Holography: Learning Accurate Wave
  Propagation Models for 3D Holographic Virtual and Augmented Reality
  Displays}.
\newblock \bibinfo{journal}{\emph{ACM Trans. Graph. (SIGGRAPH Asia)}}
  (\bibinfo{year}{2021}).
\newblock


\bibitem[\protect\citeauthoryear{Choi, Kim, Peng, and Wetzstein}{Choi
  et~al\mbox{.}}{2021b}]%
        {choi2021optimizing}
\bibfield{author}{\bibinfo{person}{Suyeon Choi}, \bibinfo{person}{Jonghyun
  Kim}, \bibinfo{person}{Yifan Peng}, {and} \bibinfo{person}{Gordon
  Wetzstein}.} \bibinfo{year}{2021}\natexlab{b}.
\newblock \showarticletitle{Optimizing image quality for holographic near-eye
  displays with michelson holography}.
\newblock \bibinfo{journal}{\emph{Optica}} \bibinfo{volume}{8},
  \bibinfo{number}{2} (\bibinfo{year}{2021}), \bibinfo{pages}{143--146}.
\newblock


\bibitem[\protect\citeauthoryear{Chung, Ahn, and Bengio}{Chung
  et~al\mbox{.}}{2016}]%
        {chung2016hierarchical}
\bibfield{author}{\bibinfo{person}{Junyoung Chung}, \bibinfo{person}{Sungjin
  Ahn}, {and} \bibinfo{person}{Yoshua Bengio}.}
  \bibinfo{year}{2016}\natexlab{}.
\newblock \showarticletitle{Hierarchical multiscale recurrent neural networks}.
\newblock \bibinfo{journal}{\emph{arXiv preprint arXiv:1609.01704}}
  (\bibinfo{year}{2016}).
\newblock


\bibitem[\protect\citeauthoryear{Fienup}{Fienup}{1982}]%
        {Fienup:1982}
\bibfield{author}{\bibinfo{person}{James~R Fienup}.}
  \bibinfo{year}{1982}\natexlab{}.
\newblock \showarticletitle{Phase retrieval algorithms: a comparison}.
\newblock \bibinfo{journal}{\emph{Applied optics}} \bibinfo{volume}{21},
  \bibinfo{number}{15} (\bibinfo{year}{1982}), \bibinfo{pages}{2758--2769}.
\newblock


\bibitem[\protect\citeauthoryear{Gerchberg}{Gerchberg}{1972}]%
        {Gerchberg:1972}
\bibfield{author}{\bibinfo{person}{Ralph~W Gerchberg}.}
  \bibinfo{year}{1972}\natexlab{}.
\newblock \showarticletitle{A practical algorithm for the determination of
  phase from image and diffraction plane pictures}.
\newblock \bibinfo{journal}{\emph{Optik}}  \bibinfo{volume}{35}
  (\bibinfo{year}{1972}), \bibinfo{pages}{237--246}.
\newblock


\bibitem[\protect\citeauthoryear{Goodman}{Goodman}{2014}]%
        {Goodman:2014}
\bibfield{author}{\bibinfo{person}{Joseph~W. Goodman}.}
  \bibinfo{year}{2014}\natexlab{}.
\newblock \showarticletitle{Holography Viewed from the Perspective of the Light
  Field Camera}. In \bibinfo{booktitle}{\emph{Fringe 2013}},
  \bibfield{editor}{\bibinfo{person}{Wolfgang Osten}} (Ed.).
  \bibinfo{publisher}{Springer Berlin Heidelberg}, \bibinfo{pages}{3--15}.
\newblock


\bibitem[\protect\citeauthoryear{Horisaki, Nishizaki, Kitaguchi, Saito, and
  Tanida}{Horisaki et~al\mbox{.}}{2021}]%
        {Horisaki:21}
\bibfield{author}{\bibinfo{person}{Ryoichi Horisaki}, \bibinfo{person}{Yohei
  Nishizaki}, \bibinfo{person}{Katsuhisa Kitaguchi}, \bibinfo{person}{Mamoru
  Saito}, {and} \bibinfo{person}{Jun Tanida}.} \bibinfo{year}{2021}\natexlab{}.
\newblock \showarticletitle{Three-dimensional deeply generated holography}.
\newblock \bibinfo{journal}{\emph{Appl. Opt.}} \bibinfo{volume}{60},
  \bibinfo{number}{4} (\bibinfo{year}{2021}), \bibinfo{pages}{A323--A328}.
\newblock


\bibitem[\protect\citeauthoryear{Horisaki, Takagi, and Tanida}{Horisaki
  et~al\mbox{.}}{2018}]%
        {Horisaki:2018}
\bibfield{author}{\bibinfo{person}{Ryoichi Horisaki}, \bibinfo{person}{Ryosuke
  Takagi}, {and} \bibinfo{person}{Jun Tanida}.}
  \bibinfo{year}{2018}\natexlab{}.
\newblock \showarticletitle{Deep-learning-generated holography}.
\newblock \bibinfo{journal}{\emph{Applied optics}} \bibinfo{volume}{57},
  \bibinfo{number}{14} (\bibinfo{year}{2018}), \bibinfo{pages}{3859--3863}.
\newblock


\bibitem[\protect\citeauthoryear{Hsueh and Sawchuk}{Hsueh and Sawchuk}{1978}]%
        {Hsueh:1978}
\bibfield{author}{\bibinfo{person}{Chung-Kai Hsueh} {and}
  \bibinfo{person}{Alexander~A. Sawchuk}.} \bibinfo{year}{1978}\natexlab{}.
\newblock \showarticletitle{Computer-generated double-phase holograms}.
\newblock \bibinfo{journal}{\emph{Applied optics}} \bibinfo{volume}{17},
  \bibinfo{number}{24} (\bibinfo{year}{1978}), \bibinfo{pages}{3874--3883}.
\newblock


\bibitem[\protect\citeauthoryear{Jang, Bang, Moon, Kim, Lee, and Lee}{Jang
  et~al\mbox{.}}{2017}]%
        {Jang:2017}
\bibfield{author}{\bibinfo{person}{Changwon Jang}, \bibinfo{person}{Kiseung
  Bang}, \bibinfo{person}{Seokil Moon}, \bibinfo{person}{Jonghyun Kim},
  \bibinfo{person}{Seungjae Lee}, {and} \bibinfo{person}{Byoungho Lee}.}
  \bibinfo{year}{2017}\natexlab{}.
\newblock \showarticletitle{Retinal 3D: augmented reality near-eye display via
  pupil-tracked light field projection on retina}.
\newblock \bibinfo{journal}{\emph{ACM Trans. Graph. (SIGGRAPH Asia)}}
  \bibinfo{volume}{36}, \bibinfo{number}{6} (\bibinfo{year}{2017}).
\newblock


\bibitem[\protect\citeauthoryear{Jang, Gu, and Poole}{Jang
  et~al\mbox{.}}{2016}]%
        {jang2016categorical}
\bibfield{author}{\bibinfo{person}{Eric Jang}, \bibinfo{person}{Shixiang Gu},
  {and} \bibinfo{person}{Ben Poole}.} \bibinfo{year}{2016}\natexlab{}.
\newblock \showarticletitle{Categorical reparameterization with
  gumbel-softmax}.
\newblock \bibinfo{journal}{\emph{arXiv preprint arXiv:1611.01144}}
  (\bibinfo{year}{2016}).
\newblock


\bibitem[\protect\citeauthoryear{Javidi, Carnicer, Anand, Barbastathis, Chen,
  Ferraro, Goodman, Horisaki, Khare, Kujawinska, Leitgeb, Marquet, Nomura,
  Ozcan, Park, Pedrini, Picart, Rosen, Saavedra, Shaked, Stern, Tajahuerce,
  Tian, Wetzstein, and Yamaguchi}{Javidi et~al\mbox{.}}{2021}]%
        {Javidi:21}
\bibfield{author}{\bibinfo{person}{Bahram Javidi}, \bibinfo{person}{Artur
  Carnicer}, \bibinfo{person}{Arun Anand}, \bibinfo{person}{George
  Barbastathis}, \bibinfo{person}{Wen Chen}, \bibinfo{person}{Pietro Ferraro},
  \bibinfo{person}{J.~W. Goodman}, \bibinfo{person}{Ryoichi Horisaki},
  \bibinfo{person}{Kedar Khare}, \bibinfo{person}{Malgorzata Kujawinska},
  \bibinfo{person}{Rainer~A. Leitgeb}, \bibinfo{person}{Pierre Marquet},
  \bibinfo{person}{Takanori Nomura}, \bibinfo{person}{Aydogan Ozcan},
  \bibinfo{person}{YongKeun Park}, \bibinfo{person}{Giancarlo Pedrini},
  \bibinfo{person}{Pascal Picart}, \bibinfo{person}{Joseph Rosen},
  \bibinfo{person}{Genaro Saavedra}, \bibinfo{person}{Natan~T. Shaked},
  \bibinfo{person}{Adrian Stern}, \bibinfo{person}{Enrique Tajahuerce},
  \bibinfo{person}{Lei Tian}, \bibinfo{person}{Gordon Wetzstein}, {and}
  \bibinfo{person}{Masahiro Yamaguchi}.} \bibinfo{year}{2021}\natexlab{}.
\newblock \showarticletitle{Roadmap on digital holography}.
\newblock \bibinfo{journal}{\emph{Opt. Express}} \bibinfo{volume}{29},
  \bibinfo{number}{22} (\bibinfo{year}{2021}).
\newblock


\bibitem[\protect\citeauthoryear{Kang, Yamaguchi, and Yoshikawa}{Kang
  et~al\mbox{.}}{2008}]%
        {Kang:08}
\bibfield{author}{\bibinfo{person}{Hoonjong Kang}, \bibinfo{person}{Takeshi
  Yamaguchi}, {and} \bibinfo{person}{Hiroshi Yoshikawa}.}
  \bibinfo{year}{2008}\natexlab{}.
\newblock \showarticletitle{Accurate phase-added stereogram to improve the
  coherent stereogram}.
\newblock \bibinfo{journal}{\emph{Appl. Opt.}} \bibinfo{volume}{47},
  \bibinfo{number}{19} (\bibinfo{year}{2008}).
\newblock


\bibitem[\protect\citeauthoryear{Kavakl{i}, Urey, and Ak\c{s}it}{Kavakl{i}
  et~al\mbox{.}}{2022}]%
        {Kavakli:22}
\bibfield{author}{\bibinfo{person}{Koray Kavakl{i}}, \bibinfo{person}{Hakan
  Urey}, {and} \bibinfo{person}{Kaan Ak\c{s}it}.}
  \bibinfo{year}{2022}\natexlab{}.
\newblock \showarticletitle{Learned holographic light transport}.
\newblock \bibinfo{journal}{\emph{Appl. Opt.}} \bibinfo{volume}{61},
  \bibinfo{number}{5} (\bibinfo{year}{2022}), \bibinfo{pages}{B50--B55}.
\newblock


\bibitem[\protect\citeauthoryear{Ketchum and Blanche}{Ketchum and
  Blanche}{2021}]%
        {ketchum2021diffraction}
\bibfield{author}{\bibinfo{person}{Remington~S Ketchum} {and}
  \bibinfo{person}{Pierre-Alexandre Blanche}.} \bibinfo{year}{2021}\natexlab{}.
\newblock \showarticletitle{Diffraction efficiency characteristics for
  MEMS-based phase-only spatial light modulator with nonlinear phase
  distribution}. In \bibinfo{booktitle}{\emph{Photonics}},
  Vol.~\bibinfo{volume}{8}. Multidisciplinary Digital Publishing Institute,
  \bibinfo{pages}{62}.
\newblock


\bibitem[\protect\citeauthoryear{Kim, Nam, Bang, Lee, Lee, Jeong, Seo, and
  Lee}{Kim et~al\mbox{.}}{2021}]%
        {Kim:21}
\bibfield{author}{\bibinfo{person}{Dongyeon Kim}, \bibinfo{person}{Seung-Woo
  Nam}, \bibinfo{person}{Kiseung Bang}, \bibinfo{person}{Byounghyo Lee},
  \bibinfo{person}{Seungjae Lee}, \bibinfo{person}{Youngmo Jeong},
  \bibinfo{person}{Jong-Mo Seo}, {and} \bibinfo{person}{Byoungho Lee}.}
  \bibinfo{year}{2021}\natexlab{}.
\newblock \showarticletitle{Vision-correcting holographic display: evaluation
  of aberration correcting hologram}.
\newblock \bibinfo{journal}{\emph{Biomed. Opt. Express}} \bibinfo{volume}{12},
  \bibinfo{number}{8} (\bibinfo{year}{2021}), \bibinfo{pages}{5179--5195}.
\newblock


\bibitem[\protect\citeauthoryear{Kim, Gopakumar, Choi, Peng, Lopes, and
  Wetzstein}{Kim et~al\mbox{.}}{2022}]%
        {kim2022holographic}
\bibfield{author}{\bibinfo{person}{Jonghyun Kim}, \bibinfo{person}{Manu
  Gopakumar}, \bibinfo{person}{Suyeon Choi}, \bibinfo{person}{Yifan Peng},
  \bibinfo{person}{Ward Lopes}, {and} \bibinfo{person}{Gordon Wetzstein}.}
  \bibinfo{year}{2022}\natexlab{}.
\newblock \showarticletitle{Holographic glasses for virtual reality}. In
  \bibinfo{booktitle}{\emph{Proceedings of the ACM SIGGRAPH}}.
\newblock


\bibitem[\protect\citeauthoryear{Lee, Kim, Lee, Chen, and Lee}{Lee
  et~al\mbox{.}}{2022}]%
        {lee2022high}
\bibfield{author}{\bibinfo{person}{Byounghyo Lee}, \bibinfo{person}{Dongyeon
  Kim}, \bibinfo{person}{Seungjae Lee}, \bibinfo{person}{Chun Chen}, {and}
  \bibinfo{person}{Byoungho Lee}.} \bibinfo{year}{2022}\natexlab{}.
\newblock \showarticletitle{High-contrast, speckle-free, true 3D holography via
  binary CGH optimization}.
\newblock \bibinfo{journal}{\emph{arXiv preprint arXiv:2201.02619}}
  (\bibinfo{year}{2022}).
\newblock


\bibitem[\protect\citeauthoryear{Lee}{Lee}{1970}]%
        {Lee:1970}
\bibfield{author}{\bibinfo{person}{Wai~Hon Lee}.}
  \bibinfo{year}{1970}\natexlab{}.
\newblock \showarticletitle{Sampled Fourier transform hologram generated by
  computer}.
\newblock \bibinfo{journal}{\emph{Applied Optics}} \bibinfo{volume}{9},
  \bibinfo{number}{3} (\bibinfo{year}{1970}), \bibinfo{pages}{639--643}.
\newblock


\bibitem[\protect\citeauthoryear{Lucente and Galyean}{Lucente and
  Galyean}{1995}]%
        {Lucente:1995}
\bibfield{author}{\bibinfo{person}{Mark Lucente} {and}
  \bibinfo{person}{Tinsley~A Galyean}.} \bibinfo{year}{1995}\natexlab{}.
\newblock \showarticletitle{Rendering interactive holographic images}. In
  \bibinfo{booktitle}{\emph{ACM SIGGRAPH}}. \bibinfo{pages}{387--394}.
\newblock


\bibitem[\protect\citeauthoryear{Maddison, Mnih, and Teh}{Maddison
  et~al\mbox{.}}{2016}]%
        {maddison2016concrete}
\bibfield{author}{\bibinfo{person}{Chris~J Maddison}, \bibinfo{person}{Andriy
  Mnih}, {and} \bibinfo{person}{Yee~Whye Teh}.}
  \bibinfo{year}{2016}\natexlab{}.
\newblock \showarticletitle{The concrete distribution: A continuous relaxation
  of discrete random variables}.
\newblock \bibinfo{journal}{\emph{arXiv preprint arXiv:1611.00712}}
  (\bibinfo{year}{2016}).
\newblock


\bibitem[\protect\citeauthoryear{Maimone, Georgiou, and Kollin}{Maimone
  et~al\mbox{.}}{2017}]%
        {Maimone:2017}
\bibfield{author}{\bibinfo{person}{Andrew Maimone}, \bibinfo{person}{Andreas
  Georgiou}, {and} \bibinfo{person}{Joel~S Kollin}.}
  \bibinfo{year}{2017}\natexlab{}.
\newblock \showarticletitle{Holographic near-eye displays for virtual and
  augmented reality}.
\newblock \bibinfo{journal}{\emph{ACM Trans. Graph. (SIGGRAPH)}}
  \bibinfo{volume}{36}, \bibinfo{number}{4} (\bibinfo{year}{2017}),
  \bibinfo{pages}{85}.
\newblock


\bibitem[\protect\citeauthoryear{Maimone and Wang}{Maimone and Wang}{2020}]%
        {Maimone:2020}
\bibfield{author}{\bibinfo{person}{Andrew Maimone} {and}
  \bibinfo{person}{Junren Wang}.} \bibinfo{year}{2020}\natexlab{}.
\newblock \showarticletitle{Holographic Optics for Thin and Lightweight Virtual
  Reality}.
\newblock \bibinfo{journal}{\emph{ACM Trans. Graph. (SIGGRAPH)}}
  \bibinfo{volume}{39}, \bibinfo{number}{4} (\bibinfo{year}{2020}).
\newblock


\bibitem[\protect\citeauthoryear{Matsushima and Nakahara}{Matsushima and
  Nakahara}{2009}]%
        {Matsushima:2009:polygon}
\bibfield{author}{\bibinfo{person}{Kyoji Matsushima} {and}
  \bibinfo{person}{Sumio Nakahara}.} \bibinfo{year}{2009}\natexlab{}.
\newblock \showarticletitle{Extremely high-definition full-parallax
  computer-generated hologram created by the polygon-based method}.
\newblock \bibinfo{journal}{\emph{Applied optics}} \bibinfo{volume}{48},
  \bibinfo{number}{34} (\bibinfo{year}{2009}), \bibinfo{pages}{H54--H63}.
\newblock


\bibitem[\protect\citeauthoryear{Padmanaban, Peng, and Wetzstein}{Padmanaban
  et~al\mbox{.}}{2019}]%
        {Padmanaban:2019}
\bibfield{author}{\bibinfo{person}{Nitish Padmanaban}, \bibinfo{person}{Yifan
  Peng}, {and} \bibinfo{person}{Gordon Wetzstein}.}
  \bibinfo{year}{2019}\natexlab{}.
\newblock \showarticletitle{Holographic Near-eye Displays Based on Overlap-add
  Stereograms}.
\newblock \bibinfo{journal}{\emph{ACM Trans. Graph.}} \bibinfo{volume}{38},
  \bibinfo{number}{6} (\bibinfo{year}{2019}).
\newblock


\bibitem[\protect\citeauthoryear{Park}{Park}{2017}]%
        {Park:2017}
\bibfield{author}{\bibinfo{person}{Jae-Hyeung Park}.}
  \bibinfo{year}{2017}\natexlab{}.
\newblock \showarticletitle{Recent progress in computer-generated holography
  for three-dimensional scenes}.
\newblock \bibinfo{journal}{\emph{Journal of Information Display}}
  \bibinfo{volume}{18}, \bibinfo{number}{1} (\bibinfo{year}{2017}),
  \bibinfo{pages}{1--12}.
\newblock


\bibitem[\protect\citeauthoryear{Peng, Choi, , Kim, and Wetzstein}{Peng
  et~al\mbox{.}}{2021}]%
        {peng2021partiallycoherent}
\bibfield{author}{\bibinfo{person}{Yifan Peng}, \bibinfo{person}{Suyeon Choi},
  \bibinfo{person}{}, \bibinfo{person}{Jonghyun Kim}, {and}
  \bibinfo{person}{Gordon Wetzstein}.} \bibinfo{year}{2021}\natexlab{}.
\newblock \showarticletitle{Speckle-free holography with partially coherent
  light sources and camera-in-the-loop calibration}.
\newblock \bibinfo{journal}{\emph{Science Advances}} (\bibinfo{year}{2021}).
\newblock


\bibitem[\protect\citeauthoryear{Peng, Choi, Padmanaban, and Wetzstein}{Peng
  et~al\mbox{.}}{2020}]%
        {peng2020neural}
\bibfield{author}{\bibinfo{person}{Yifan Peng}, \bibinfo{person}{Suyeon Choi},
  \bibinfo{person}{Nitish Padmanaban}, {and} \bibinfo{person}{Gordon
  Wetzstein}.} \bibinfo{year}{2020}\natexlab{}.
\newblock \showarticletitle{Neural holography with camera-in-the-loop
  training}.
\newblock \bibinfo{journal}{\emph{ACM Trans. Graph.}} \bibinfo{volume}{39},
  \bibinfo{number}{6} (\bibinfo{year}{2020}), \bibinfo{pages}{1--14}.
\newblock


\bibitem[\protect\citeauthoryear{Shi, Huang, Lopes, Matusik, and Luebke}{Shi
  et~al\mbox{.}}{2017}]%
        {Shi:2017}
\bibfield{author}{\bibinfo{person}{Liang Shi}, \bibinfo{person}{Fu-Chung
  Huang}, \bibinfo{person}{Ward Lopes}, \bibinfo{person}{Wojciech Matusik},
  {and} \bibinfo{person}{David Luebke}.} \bibinfo{year}{2017}\natexlab{}.
\newblock \showarticletitle{Near-eye Light Field Holographic Rendering with
  Spherical Waves for Wide Field of View Interactive 3D Computer Graphics}.
\newblock \bibinfo{journal}{\emph{ACM Trans. Graph.}} \bibinfo{volume}{36},
  \bibinfo{number}{6} (\bibinfo{year}{2017}).
\newblock


\bibitem[\protect\citeauthoryear{Shi, Li, Kim, Kellnhofer, and Matusik}{Shi
  et~al\mbox{.}}{2021}]%
        {shi2021towards}
\bibfield{author}{\bibinfo{person}{Liang Shi}, \bibinfo{person}{Beichen Li},
  \bibinfo{person}{Changil Kim}, \bibinfo{person}{Petr Kellnhofer}, {and}
  \bibinfo{person}{Wojciech Matusik}.} \bibinfo{year}{2021}\natexlab{}.
\newblock \showarticletitle{Towards real-time photorealistic 3D holography with
  deep neural networks}.
\newblock \bibinfo{journal}{\emph{Nature}} \bibinfo{volume}{591},
  \bibinfo{number}{7849} (\bibinfo{year}{2021}), \bibinfo{pages}{234--239}.
\newblock


\bibitem[\protect\citeauthoryear{Wakunami, Yamashita, and Yamaguchi}{Wakunami
  et~al\mbox{.}}{2013}]%
        {Wakunami:2013}
\bibfield{author}{\bibinfo{person}{Koki Wakunami}, \bibinfo{person}{Hiroaki
  Yamashita}, {and} \bibinfo{person}{Masahiro Yamaguchi}.}
  \bibinfo{year}{2013}\natexlab{}.
\newblock \showarticletitle{Occlusion culling for computer generated hologram
  based on ray-wavefront conversion}.
\newblock \bibinfo{journal}{\emph{Optics express}} \bibinfo{volume}{21},
  \bibinfo{number}{19} (\bibinfo{year}{2013}), \bibinfo{pages}{21811--21822}.
\newblock


\bibitem[\protect\citeauthoryear{Yaras, Kang, and Onural}{Yaras
  et~al\mbox{.}}{2010}]%
        {Yaras:2010}
\bibfield{author}{\bibinfo{person}{Fahri Yaras}, \bibinfo{person}{Hoonjong
  Kang}, {and} \bibinfo{person}{Levent Onural}.}
  \bibinfo{year}{2010}\natexlab{}.
\newblock \showarticletitle{State of the Art in Holographic Displays: A
  Survey}.
\newblock \bibinfo{journal}{\emph{Journal of Display Technology}}
  \bibinfo{volume}{6}, \bibinfo{number}{10} (\bibinfo{year}{2010}),
  \bibinfo{pages}{443--454}.
\newblock


\bibitem[\protect\citeauthoryear{Yoo, Jo, Nam, Chen, and Lee}{Yoo
  et~al\mbox{.}}{2021}]%
        {Yoo:21}
\bibfield{author}{\bibinfo{person}{Dongheon Yoo}, \bibinfo{person}{Youngjin
  Jo}, \bibinfo{person}{Seung-Woo Nam}, \bibinfo{person}{Chun Chen}, {and}
  \bibinfo{person}{Byoungho Lee}.} \bibinfo{year}{2021}\natexlab{}.
\newblock \showarticletitle{Optimization of computer-generated holograms
  featuring phase randomness control}.
\newblock \bibinfo{journal}{\emph{Opt. Lett.}} \bibinfo{volume}{46},
  \bibinfo{number}{19} (\bibinfo{year}{2021}), \bibinfo{pages}{4769--4772}.
\newblock


\bibitem[\protect\citeauthoryear{Zenke and Ganguli}{Zenke and Ganguli}{2018}]%
        {zenke2018superspike}
\bibfield{author}{\bibinfo{person}{Friedemann Zenke} {and}
  \bibinfo{person}{Surya Ganguli}.} \bibinfo{year}{2018}\natexlab{}.
\newblock \showarticletitle{Superspike: Supervised learning in multilayer
  spiking neural networks}.
\newblock \bibinfo{journal}{\emph{Neural computation}} \bibinfo{volume}{30},
  \bibinfo{number}{6} (\bibinfo{year}{2018}), \bibinfo{pages}{1514--1541}.
\newblock


\bibitem[\protect\citeauthoryear{Zhang, Cao, and Jin}{Zhang
  et~al\mbox{.}}{2017}]%
        {Zhang:2017}
\bibfield{author}{\bibinfo{person}{Hao Zhang}, \bibinfo{person}{Liangcai Cao},
  {and} \bibinfo{person}{Guofan Jin}.} \bibinfo{year}{2017}\natexlab{}.
\newblock \showarticletitle{Computer-generated hologram with occlusion effect
  using layer-based processing}.
\newblock \bibinfo{journal}{\emph{Applied optics}} \bibinfo{volume}{56},
  \bibinfo{number}{13} (\bibinfo{year}{2017}).
\newblock


\bibitem[\protect\citeauthoryear{Zhang, Collings, Chen, Crossland, Chu, and
  Xie}{Zhang et~al\mbox{.}}{2011}]%
        {Zhang:2011}
\bibfield{author}{\bibinfo{person}{Hao Zhang}, \bibinfo{person}{Neil Collings},
  \bibinfo{person}{Jing Chen}, \bibinfo{person}{Bill~A Crossland},
  \bibinfo{person}{Daping Chu}, {and} \bibinfo{person}{Jinghui Xie}.}
  \bibinfo{year}{2011}\natexlab{}.
\newblock \showarticletitle{Full parallax three-dimensional display with
  occlusion effect using computer generated hologram}.
\newblock \bibinfo{journal}{\emph{Optical Engineering}} \bibinfo{volume}{50},
  \bibinfo{number}{7} (\bibinfo{year}{2011}), \bibinfo{pages}{074003}.
\newblock


\bibitem[\protect\citeauthoryear{Zhang and Levoy}{Zhang and Levoy}{2009}]%
        {Zhang:2009}
\bibfield{author}{\bibinfo{person}{Zhengyun Zhang} {and} \bibinfo{person}{M.
  Levoy}.} \bibinfo{year}{2009}\natexlab{}.
\newblock \showarticletitle{Wigner distributions and how they relate to the
  light field}. In \bibinfo{booktitle}{\emph{Proc. ICCP}}.
  \bibinfo{publisher}{IEEE}, \bibinfo{pages}{1--10}.
\newblock


\bibitem[\protect\citeauthoryear{Ziegler, Bucheli, Ahrenberg, Magnor, and
  Gross}{Ziegler et~al\mbox{.}}{2007}]%
        {Ziegler:2007}
\bibfield{author}{\bibinfo{person}{Remo Ziegler}, \bibinfo{person}{Simon
  Bucheli}, \bibinfo{person}{Lukas Ahrenberg}, \bibinfo{person}{Marcus Magnor},
  {and} \bibinfo{person}{Markus Gross}.} \bibinfo{year}{2007}\natexlab{}.
\newblock \showarticletitle{A Bidirectional Light Field-Hologram Transform}. In
  \bibinfo{booktitle}{\emph{Computer Graphics Forum (Eurographics)}},
  Vol.~\bibinfo{volume}{26}. \bibinfo{pages}{435--446}.
\newblock


\end{thebibliography}



\begin{thebibliography}{9}


\ifx \showCODEN    \undefined \def \showCODEN     #1{\unskip}     \fi
\ifx \showDOI      \undefined \def \showDOI       #1{#1}\fi
\ifx \showISBNx    \undefined \def \showISBNx     #1{\unskip}     \fi
\ifx \showISBNxiii \undefined \def \showISBNxiii  #1{\unskip}     \fi
\ifx \showISSN     \undefined \def \showISSN      #1{\unskip}     \fi
\ifx \showLCCN     \undefined \def \showLCCN      #1{\unskip}     \fi
\ifx \shownote     \undefined \def \shownote      #1{#1}          \fi
\ifx \showarticletitle \undefined \def \showarticletitle #1{#1}   \fi
\ifx \showURL      \undefined \def \showURL       {\relax}        \fi
\providecommand\bibfield[2]{#2}
\providecommand\bibinfo[2]{#2}
\providecommand\natexlab[1]{#1}
\providecommand\showeprint[2][]{arXiv:#2}

\bibitem[\protect\citeauthoryear{Choi, Gopakumar, Peng, Kim, and
  Wetzstein}{Choi et~al\mbox{.}}{2021}]%
        {choi2021neural3d}
\bibfield{author}{\bibinfo{person}{Suyeon Choi}, \bibinfo{person}{Manu
  Gopakumar}, \bibinfo{person}{Yifan Peng}, \bibinfo{person}{Jonghyun Kim},
  {and} \bibinfo{person}{Gordon Wetzstein}.} \bibinfo{year}{2021}\natexlab{}.
\newblock \showarticletitle{Neural 3D Holography: Learning Accurate Wave
  Propagation Models for 3D Holographic Virtual and Augmented Reality
  Displays}.
\newblock \bibinfo{journal}{\emph{ACM Trans. Graph. (SIGGRAPH Asia)}}
  (\bibinfo{year}{2021}).
\newblock


\bibitem[\protect\citeauthoryear{Goodman}{Goodman}{2005}]%
        {Goodman:2005}
\bibfield{author}{\bibinfo{person}{Joseph~W Goodman}.}
  \bibinfo{year}{2005}\natexlab{}.
\newblock \bibinfo{booktitle}{\emph{Introduction to Fourier optics}}.
\newblock \bibinfo{publisher}{Roberts and Company}.
\newblock


\bibitem[\protect\citeauthoryear{Jang, Gu, and Poole}{Jang
  et~al\mbox{.}}{2016}]%
        {jang2016categorical}
\bibfield{author}{\bibinfo{person}{Eric Jang}, \bibinfo{person}{Shixiang Gu},
  {and} \bibinfo{person}{Ben Poole}.} \bibinfo{year}{2016}\natexlab{}.
\newblock \showarticletitle{Categorical reparameterization with
  gumbel-softmax}.
\newblock \bibinfo{journal}{\emph{arXiv preprint arXiv:1611.01144}}
  (\bibinfo{year}{2016}).
\newblock


\bibitem[\protect\citeauthoryear{Lee, Kim, Lee, Chen, and Lee}{Lee
  et~al\mbox{.}}{2022}]%
        {lee2022high}
\bibfield{author}{\bibinfo{person}{Byounghyo Lee}, \bibinfo{person}{Dongyeon
  Kim}, \bibinfo{person}{Seungjae Lee}, \bibinfo{person}{Chun Chen}, {and}
  \bibinfo{person}{Byoungho Lee}.} \bibinfo{year}{2022}\natexlab{}.
\newblock \showarticletitle{High-contrast, speckle-free, true 3D holography via
  binary CGH optimization}.
\newblock \bibinfo{journal}{\emph{arXiv preprint arXiv:2201.02619}}
  (\bibinfo{year}{2022}).
\newblock


\bibitem[\protect\citeauthoryear{Maddison, Mnih, and Teh}{Maddison
  et~al\mbox{.}}{2016}]%
        {maddison2016concrete}
\bibfield{author}{\bibinfo{person}{Chris~J Maddison}, \bibinfo{person}{Andriy
  Mnih}, {and} \bibinfo{person}{Yee~Whye Teh}.}
  \bibinfo{year}{2016}\natexlab{}.
\newblock \showarticletitle{The concrete distribution: A continuous relaxation
  of discrete random variables}.
\newblock \bibinfo{journal}{\emph{arXiv preprint arXiv:1611.00712}}
  (\bibinfo{year}{2016}).
\newblock


\bibitem[\protect\citeauthoryear{Padmanaban, Peng, and Wetzstein}{Padmanaban
  et~al\mbox{.}}{2019}]%
        {Padmanaban:2019}
\bibfield{author}{\bibinfo{person}{Nitish Padmanaban}, \bibinfo{person}{Yifan
  Peng}, {and} \bibinfo{person}{Gordon Wetzstein}.}
  \bibinfo{year}{2019}\natexlab{}.
\newblock \showarticletitle{Holographic Near-eye Displays Based on Overlap-add
  Stereograms}.
\newblock \bibinfo{journal}{\emph{ACM Trans. Graph.}} \bibinfo{volume}{38},
  \bibinfo{number}{6} (\bibinfo{year}{2019}).
\newblock


\bibitem[\protect\citeauthoryear{Peng, Choi, Padmanaban, and Wetzstein}{Peng
  et~al\mbox{.}}{2020}]%
        {peng2020neural}
\bibfield{author}{\bibinfo{person}{Yifan Peng}, \bibinfo{person}{Suyeon Choi},
  \bibinfo{person}{Nitish Padmanaban}, {and} \bibinfo{person}{Gordon
  Wetzstein}.} \bibinfo{year}{2020}\natexlab{}.
\newblock \showarticletitle{Neural holography with camera-in-the-loop
  training}.
\newblock \bibinfo{journal}{\emph{ACM Trans. Graph.}} \bibinfo{volume}{39},
  \bibinfo{number}{6} (\bibinfo{year}{2020}), \bibinfo{pages}{1--14}.
\newblock


\bibitem[\protect\citeauthoryear{Shi, Li, Kim, Kellnhofer, and Matusik}{Shi
  et~al\mbox{.}}{2021}]%
        {shi2021towards}
\bibfield{author}{\bibinfo{person}{Liang Shi}, \bibinfo{person}{Beichen Li},
  \bibinfo{person}{Changil Kim}, \bibinfo{person}{Petr Kellnhofer}, {and}
  \bibinfo{person}{Wojciech Matusik}.} \bibinfo{year}{2021}\natexlab{}.
\newblock \showarticletitle{Towards real-time photorealistic 3D holography with
  deep neural networks}.
\newblock \bibinfo{journal}{\emph{Nature}} \bibinfo{volume}{591},
  \bibinfo{number}{7849} (\bibinfo{year}{2021}), \bibinfo{pages}{234--239}.
\newblock


\bibitem[\protect\citeauthoryear{Yoo, Jo, Nam, Chen, and Lee}{Yoo
  et~al\mbox{.}}{2021}]%
        {Yoo:21}
\bibfield{author}{\bibinfo{person}{Dongheon Yoo}, \bibinfo{person}{Youngjin
  Jo}, \bibinfo{person}{Seung-Woo Nam}, \bibinfo{person}{Chun Chen}, {and}
  \bibinfo{person}{Byoungho Lee}.} \bibinfo{year}{2021}\natexlab{}.
\newblock \showarticletitle{Optimization of computer-generated holograms
  featuring phase randomness control}.
\newblock \bibinfo{journal}{\emph{Opt. Lett.}} \bibinfo{volume}{46},
  \bibinfo{number}{19} (\bibinfo{year}{2021}), \bibinfo{pages}{4769--4772}.
\newblock


\end{thebibliography}
